\documentclass[a4paper,fleqn,usenatbib]{mnras}

\usepackage[T1]{fontenc}
\usepackage{ae,aecompl}

\usepackage{relsize}

\usepackage{graphicx}	
\usepackage{amsmath}	
\usepackage{amssymb}	

\usepackage{enumitem}

\usepackage{booktabs}
\usepackage{longtable}
\usepackage{array}
\usepackage{pdflscape}
\usepackage{tabularx}
\usepackage{rotating}

\usepackage{subcaption}
\usepackage{color}

\newcommand{\eqb}{\begin{equation}}
\newcommand{\eqe}{\end{equation}}
\newcommand{\dmb}{\begin{displaymath}}
\newcommand{\dme}{\end{displaymath}}

\newcommand{\eab}{\begin{eqnarray}}
\newcommand{\eae}{\end{eqnarray}}

\newcommand{\be}{\begin{equation}}
\newcommand{\ee}{\end{equation}}

\newcommand{\LC}{\textnormal{\tiny \textsc{l}}}
\newcommand{\SU}{\textnormal{\tiny \textsc{ym}}}
\newcommand{\YM}{\textnormal{\tiny \textsc{ym}}}
\newcommand{\CMB}{\textnormal{\tiny \textsc{cmb}}}

\newcommand{\CMBbold}{\text{\tiny \textbf{\textsc{CMB}}}}

\newcolumntype{C}{>{$\displaystyle} c <{$}}

\title[SU(2)$_{\CMB}$ and the cosmological model]{
SU(2)$_\CMBbold$ and the cosmological model:\\ angular power spectra}

\author[Hahn, Hofmann, \& Kramer]{
Steffen Hahn$^{1}$,
Ralf Hofmann$^{2}$, 
Daniel Kramer$^{3}$
\\
$^{1}$Karlsruher Institut f\"ur Technologie, Campus Nord,  Institut f\"ur Kernphysik, 
Hermann-von-Helmholtz-Platz 1,\\ \ \ D-76344 Eggenstein-Leopoldshafen, Germany \\
$^{2}$Universit\"at Heidelberg, Institut f\"ur Theoretische Physik, Philosophenweg 16, D-69120 Heidelberg, Germany\\ 
$^{3}$Karlsruher Institut f\"ur Technologie, Campus S\"ud, Kaiserstr. 12,  D-76131 Karlsruhe, Germany \\
}

\date{Accepted XXX. Received YYY; in original form ZZZ}

\pubyear{2018}

\begin{document}
\label{firstpage}
\pagerange{\pageref{firstpage}--\pageref{lastpage}}
\maketitle

\begin{abstract}
Driven by the CMB temperature-redshift ($T$-$z$) relation as demanded by deconfining SU(2) Yang-Mills thermodynamics, an according cosmological model is proposed and analysed. 
This model -- SU(2)$_\CMB$ -- exhibits a dark sector, representing $\Lambda$CDM with a certain late-time dark-matter density 
which transitions to a reduced (present-day) 
density parameter at high $z$. We statistically analyse constraints on cosmological parameters directly imposed 
by the values of the standard co-moving ruler $r_d$ and the angular size of the sound 
horizon $\theta_*$. Compared to the $\Lambda$CDM best fit to 2015 Planck data,
we require an increased (present-day) dark matter density when $r_d\cdot H_0=$\,const and a 
value $H_0\sim 73.5$ km\,s$^{-1}$Mpc$^{-1}$ -- typical for local extractions -- are used. The ratio between the 
density parameters of primordial and late-time dark matter ranges between 0.5 
and 0.7. We confirm this trend by fitting the predictions of SU(2)$_\CMB$, obtained from a modified CLASS code, to the angular power spectra TT, TE, and EE. We consider adiabatic, scalar primordial curvature perturbations and distinguish two treatments of thermal quasi-particles in the perturbation equations. Best fits predict a red-tilted primordial power spectrum. Moreover, a low baryon density is obtained compared with the coincidence value of BBN, the $\Lambda$CDM best fit of the 2015 Planck data, and the observed deuterium abundance. Our derived values of $H_0$ support the results of local cosmological observations. Also, there is a tendency for late reionisation.
\textcolor{white}{Pdflatex could not properly compile abstracts with less than 1700 characters?!????????????}
\end{abstract}





\section{Introduction}

The last two and a half decades have converted cos\-mo\-lo\-gy into a pre\-cision science, 
resulting in independent data sets which 
constrain models of our universe: 
(i) large-scale structure surveys of the matter correlation function, suggesting the existence of a standard 
ruler $r_d$ set by the physics of baryonic acoustic oscillations, 
e.g., \cite{Abazajian2003,Adelman-McCarthy2008}, (ii) observations of the temperature and polarisation fluctuations in the CMB  \cite{Mather1990,Hinshaw2013} with a high angular resolution \cite{Ade2014a,Ade2016} by satellite based missions, and (iii) use of calibrated SNe Ia in distance-redshift surveys, ultimately 
changing the paradigm on the rate of late-time (low-$z$) expansion \cite{Perlmutter1998,Riess1998}. 

The $\Lambda$CDM concordance model fits most of the 
cosmological data in a satisfactory way. In particular, the accelerated expansion of 
the universe at late times due to the dominance of dark energy as well as the spatial flatness of the universe \cite{Boomerang}
can be considered secured facts. However, recent results reveal a 
considerable tension between the high value of the present expansion rate $H_0=73.48\pm 1.66$ km\,s$^{-1}$Mpc$^{-1}$ extracted 
from calibrated distances to SNe Ia (local observation \cite{Riess2018,Cardona2016}) and the low values $H_0=66.93\pm 0.62$ km\,s$^{-1}$Mpc$^{-1}$ obtained from the 2015 Planck data or $H_0=69.1^{+0.4}_{-0.6}$ km\,s$^{-1}$Mpc$^{-1}$ from the clustering of galaxies (global observations \cite{Adghanim2016II, DES2017}). An independent distance estimator, appealing to time delays from gravitational lensing, extracts a high value of $H_0=72.8\pm 2.4$ km\,s$^{-1}$Mpc$^{-1}$ 
(local observation \cite{Bonvin2017}). The discrepancy between the local and the global values of $H_0$ is unlikely 
to be resolved by sample variance, local matter-density fluctuations, or a directional bias in 
SNe Ia observations \cite{Marra2013,Odderskov2014,Odderskov2017,WuHuterer2017}: their effect $\Delta H_0\sim 0.31$ km\,s$^{-1}$Mpc$^{-1}$ is much smaller than the local-global discrepancy
$\Delta H_0 \sim 6$ km\,s$^{-1}$Mpc$^{-1}$. 

In \cite{HH2017} a modification 
of the high-redshift cosmological model, dubbed SU(2)$_\CMB$, was proposed. In this model the U(1) gauge 
group of electromagnetism is replaced by an SU(2) gauge principle (Yang-Mills theory), motivated theoretically in \cite{Hofmann2016a}
and observationally by an excess of CMB line temperatures at low frequencies, see \cite{Fixsen2011,Hofmann2009} 
and references therein. The work \cite{HH2017} mainly was concerned with high-$z$ implications of this 
model for the co-moving sound horizon at baryon drag $r_d$, in turn 
fixing the value of $H_0$ \cite{Bernal2016}. 
In analysing this model further towards its implications for the CMB 
angular power spectra, the present authors have noticed a number of 
oversights in \cite{HH2017}. Namely, the parameter $R_{{\rm SU(2)}_\CMB}$, governing the sound velocity in 
the baryon-Yang-Mills plasma, was defined conventionally which violates 
energy-conservation in the extended situation. 
Also, the dark-sector model turns out to be too radical in excluding 
primordial dark matter all together. In the present work these shortcomings 
are eliminated. We now demonstrate SU(2)$_\CMB$'s potential to resolve the tension in $H_0$ between local cosmology and 
$\Lambda$CDM fits to the 2015 Planck data. Our present findings favour a late reionisation as suggested by observation of the Gunn-Peterson trough in the spectra of distant quasars \cite{Becker2001}. However, these encouraging results are at the expense of admitting a red-tilted spectrum of primordial adiabatic, scalar curvature perturbations, a high 
matter density, and a low baryon density.      

This work is organised as follows. In Sec.\,\ref{CM} we introduce the 
cosmological model SU(2)$_\CMB$. The pronounced modification of the dark sector in this 
model is driven by an unconventional $T$-$z$ relation in deconfining SU(2) Yang-Mills thermodynamics, 
assuming that the CMB is described by such a theory. In addition, the conversion 
between neutrino and CMB temperature is changed. We also briefly review
linear cosmological perturbation theory in conformal Newtonian gauge with a special focus on issues 
arising in SU(2)$_\CMB$ due to the emergence of a non-primordial dark-matter component and the additional gauge-mode degrees of freedom. To constrain the cosmological parameters we perform a statistical analysis in Sec.\,\ref{SHBDtheta} by imposing the values of the angular size of the sound horizon $\theta_*$ and the standard co-moving ruler $r_d$. It turns out that these constraints only determine the total late-time and the primordial dark-matter densities while other parameters remain unfixed. 
In Sec.\,\ref{FO} computations of the angular power spectra of TT, TE, and EE as well as likelihood maximisations w.r.t. the 2015 Planck data are carried out after a short discussion of the modifications in CLASS modules. As a result, we obtain best fits (minima of co-profiles) which predict a red-tilted power spectrum of primordial, adiabatic, and scalar curvature perturbations, a low baryon density, and a high total-matter as well as dark-matter density at late times. We see good agreement of $H_0$ with local cosmological observation, and a tendency for a low redshift of reionisation $z_{\rm re}$, as extracted from spectral 
observations of distant quasars. These results are discussed in view of independent 
constraints. Finally, in Sec.\,\ref{SD} we summarize our results and sketch how the currently 
missing treatment of radiative effects at low $z$ could be implemented. We also emphasise the necessity of understanding the ``microscopics" in the purported depercolation physics of the modified dark sector.    

\section{Modified cosmological model\label{CM}}

Let us start by discussing the theoretical basis 
for SU(2)$_\CMB$. For a detailed treatise of deconfining 
SU(2) Yang-Mills thermodynamics the reader may consult Ch.\,5 of 
\cite{Hofmann2016a}, \cite{Hofmann2009} for the observational fixation of the Yang-Mills scale 
(or critical temperature $T_c$), and \cite{Hofmann2015,HH2018} for the derivation of the 
CMB-temperature--redshift ($T$-$z$) relation. For the compatibility of this $T$-$z$ relation with 
existing data, such as the thermal Sunyaev-Zeldovich effect, see \cite{HH2017}. 

We first introduce the cosmological model for the background evolution, assuming a spatially 
flat Friedmann-Lema\^itre-Robertson-Walker (FLRW) universe, also in regard to the computation 
of certain derived quantities like the co-moving sound horizon. 
This model comprises a modified dark sector which connects 
$\Lambda$CDM at low $z$ to a high-$z$ model. In addition, SU(2)$_\CMB$ demands a 
modified $T$-$z$ relation as well as a change in the conversion between neutrino 
temperature $T_\nu$ and $T$. When addressing the first-order cosmological 
perturbations in SU(2)$_\CMB$ we point out peculiarities 
surfacing in the Einstein-, Euler-, and Boltzmann equations. 
Moreover, the assumed (instantaneous) emergence of dark matter via the depercolation of 
Planck-scale axion vortices requires a modelling of the initial conditions with respect to its 
density contrast and its divergence of fluid velocity \cite{MB1995}.   

\subsection{Background model}

\subsubsection{Modified temperature-redshift relation}

It is worthwhile to repeat the discussion of the $T$-$z$ relation in 
SU(2)$_\CMB$, see also \cite{Hofmann2015,HH2018}, because 
it implies changes for the dark-matter content at high $z$ \cite{HH2017}. 

\begin{figure}
\centering
\includegraphics[width=\columnwidth]{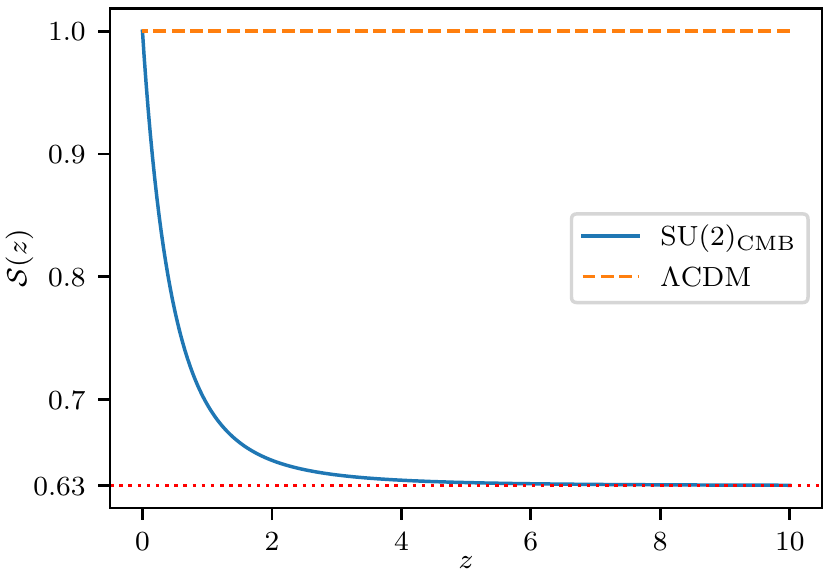}
\caption{\protect{\label{Fig-1}} The function ${\cal S}(z)$ of Eq.\,(\ref{devlinS}) which indicates 
the (multiplicative) deviation from the linear $T$-$z$ relation of 
Eq.\,(\ref{TzT>>T0}). The curvature at 
low $z$ is due to the breaking of scale invariance in the deconfining SU(2) Yang-Mills plasma for $T\sim T_0=T_c$. Notice the rapid approach towards the 
asymptotics $\left(\frac{1}{4}\right)^{1/3}\approx 0.63$ with increasing $z$.}  
   
\end{figure}

Demanding energy conservation in an FLRW 
universe the following equation is to be considered
\eqb
\label{enercon}
\frac{\mbox{d}\rho_\SU}{\mbox{d}a}=-\frac{3}{a}\left(\rho_\SU+P_\SU\right)\,,
\eqe
where $\rho_\SU$ and $P_\SU$ denote energy density and pressure, respectively, in the deconfining 
phase of SU(2) Yang-Mills thermodynamics (subscript \textsc{ym}). For later use, we associate the subscript \textsc{l} with the $\Lambda$CDM model, subject to conventional U(1) photon physics. Moreover, $a$ refers to 
the cosmological scale factor, normalised such that today $a(T_0)=1$. Here, $T_c=T_0=2.725\,$K \cite{Hofmann2009} indicates the present baseline temperature of the CMB \cite{Mather1990}. Solving for $a$, the solution to 
Eq.\,(\ref{enercon}) reads
\begin{equation}\label{solt>t0}
a \equiv \frac{1}{z+1}=\exp\left(-\frac{1}{3}\log \left(\frac{s_\SU(T)}{s_\SU(T_0)}\right) \right)\,,
\end{equation}
where the entropy density $s_\SU$ is defined by 
\eqb
\label{entropydens}
s_\SU\equiv\frac{\rho_\SU+P_\SU}{T}\,.
\eqe
For $T\gg T_0$ Eq.\,(\ref{solt>t0}) simplifies \cite{Hofmann2016a} to  
\eqb
\label{TzT>>T0}
T=\left(\frac{1}{4}\right)^{1/3}T_0(z+1)\approx 0.63\,T_0(z+1)\,.
\eqe
For arbitrary $T\ge T_0$ we define the multiplicative deviation $S(z)$ 
from linear scaling as
\eqb
\label{devlinS}
{\cal S}(z)=\left(\frac{\rho_\SU(z=0)+P_\SU(z=0)}{\rho_\SU(z)+P_\SU(z)}\frac{T^4(z)}{T^4_0}\right)^{1/3}\,,
\eqe 
such that
\eqb
\label{TzT}
T={\cal S}(z)\,T_0(z+1)\,.
\eqe
Fig.\,\ref{Fig-1} depicts function ${\cal S}(z)$.             
When the contribution of the 
thermal ground state in the deconfining SU(2) plasma is negligible and all eight gauge modes\footnote{There are two polarisations for the massless mode and three polarisations for each of the two massive modes.} are ultra-relativistic, the $z$ dependence of the Yang-Mills plasma energy density $\rho_{{\SU}}$ can, by virtue of Eq.\,(\ref{TzT>>T0}), be expressed as
\eqb
\label{rhosu2}
\rho_{{\SU}}(z)=4\,\left(\frac{1}{4}\right)^{4/3}\rho_\gamma(z)=
\left(\frac{1}{4}\right)^{1/3}\rho_\gamma(z)\quad(z\gg 1)\,.
\eqe             
Here, $\rho_\gamma$ denotes the energy density of a thermal photon gas, using the $T$-$z$ relation $T=T_0(z+1)$.

\subsubsection{Modified dark sector}

\begin{figure}
\centering
\includegraphics[width=\columnwidth]{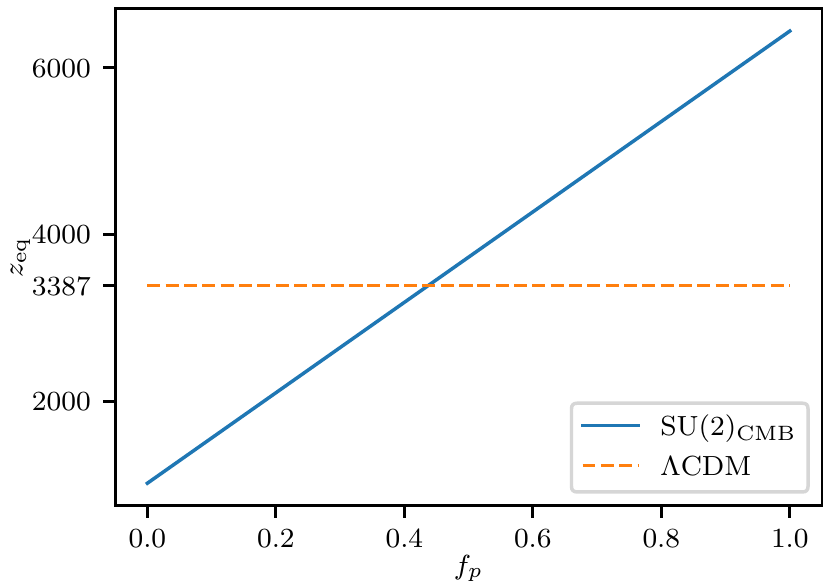}
\caption{The redshift of matter-radiation equality $z_{\rm eq}$ as a function of the 
fraction $f_p$ of primordial dark matter $\Omega_{\rm pdm,0}$ 
to today's dark matter $\Omega_\text{cdm,0}=\Omega_{\rm pdm,0}+\Omega_{\rm edm,0}$ where 
$\Omega_{\rm edm,0}$ refers to the density parameter associated with the dark-matter portion that emerges at redshift $z_p<z_\text{eq}$ (solid line). The dashed line corresponds to $z_{\rm eq}$ in $\Lambda {\rm CDM}$. \protect{\label{Fig-2}}}      
\end{figure}

Let us explore qualitatively the implications of Eq.\,(\ref{TzT>>T0}) for SU(2)$_\CMB$ at high $z$. Approximating the recombination physics by thermodynamics, 
the Thomson scattering rate $\Gamma$ can be considered a function of the recombination temperature 
$T_{\rm rec}$ only: $\Gamma=\Gamma(T_{\rm rec})$. On the other hand, 
the Hubble parameter depends on $T_{\rm rec}$ via $z_{\rm rec}$: 
$H(z_{\rm rec})=H(z(T_{\rm rec}))$. If we assume in addition 
that $H$ is matter dominated\footnote{When we speak of ``matter'' we always refer to non-relativistic, cold matter in the following.} during recombination, appealing 
to Eq.\,(\ref{TzT>>T0}), and considering that $T_{\rm rec}$ is independent of the 
cosmological model, we may formulate the decoupling condition $H=\Gamma$ as 
\eqb
\label{deccond}
H_\SU\left(z_{\SU,{\rm rec}}\right)=
H_\LC\left(z_{\LC,{\rm rec}}\right)\,.
\eqe 
At the same temperature $T$, we have 
\eqb
\label{convz}
z_{\LC}=\left(\frac{1}{4}\right)^{1/3}z_{\SU}\,.
\eqe
Taking $z_{{\LC},{\rm rec}}=1090$ \cite{Ade2016} Eq.\,(\ref{convz}) predicts 
\eqb
\label{zeqsu2}
z_{{\SU},{\rm rec}}=1730\,.
\eqe
Matter domination is stated as 
\eqb
\label{hzmatt}
H^2(z)=H_0^2\,\Omega_{m,0}(z+1)^3\,,
\eqe 
where $\Omega_{m,0}$ is the ratio of today's energy density in matter to the 
critical energy density, and $H_0$ is assumed equal in both models. Let us at first not consider a possible emergence 
of matter at any value of $z$. However, as we shall see shortly, such an emergence of matter in the dark sector at some intermediate redshift $z_p$ is implied in SU(2)$_\CMB$. Combining Eqs.\,(\ref{deccond}), (\ref{convz}), and (\ref{hzmatt}), we obtain 
\eqb
\label{Omegacon}
\Omega_{\LC,m,0} \approx 4\,\Omega_{{\SU},m,0}\,.
\eqe
Truly matter dominated recombination is not realistic, see Fig.\,\ref{Fig-2}, yet this argument
 catches an important feature of SU(2)$_\CMB$. That is, at a given redshift $z$ in some vicinity of $z_{\SU,{\rm rec}}\sim 1700$ this model's matter-density parameter $\Omega_{\SU,m,0}$ in $\Omega_{\SU,m}(z)=\Omega_{\SU,m,0}(z+1)^3$ should be suppressed 
compared to its low-$z$ value. 

When describing matter density at all $z$, 
the low value of $\Omega_{\SU,m,0}$ in Eq.\,(\ref{Omegacon}) is unacceptable in view of the successes of $\Lambda \text{CDM}$ as a low-$z$ model. Keeping in mind that the baryonic matter fraction of the today's total matter content is small in $\Lambda \text{CDM}$, we posit the emergence 
of dark matter (edm) from a dark-energy like component in the dark sector (ds)
at $z_p<z_{\SU,{\rm rec}}$ as
\begin{equation}
\begin{split}
\label{edmdef}
\Omega_{\rm ds} (z) =\, &\Omega_{\Lambda} + \Omega_{\rm pdm,0} (z+1)^3 + \\ 
&\Omega_{\rm edm,0} \left\{  \begin{array}{lr}
\left(z_{\phantom{p}} + 1\right)^3\,,&  z < z_p\\
\left(z_p + 1\right)^3 \,, & z \geq z_p
\end{array} \right.\,.  
\end{split}
\end{equation}
Here $\Omega_{\Lambda}$ and $\Omega_{\rm pdm,0}+\Omega_{\rm edm,0}\equiv \Omega_{\rm cdm,0}$ represent today's density parameters for dark energy and dark matter, respectively, $\Omega_{\rm pdm,0}$ refers to primordial dark matter for all $z$ and $\Omega_{\rm edm,0}$ to emergent dark matter for $z<z_p$. One may question the assumption of an instantaneous release of dark 
matter from dark energy as described by Eq.\,(\ref{edmdef}). In the present work, the use of model (\ref{edmdef}) is motivated by  technical simplicity. An example for such a dark sector could 
be an abundance of (non-topological) vortices and antivortices in a Planck-scale axion field (PSA) \cite{Adler1969,AdlerBardeen1969,Bell1969,Fujikawa1979,Fujikawa1980,Frieman1995,Neubert}, released by (non-thermal) Hagedorn transitions in the very early universe when Yang-Mills theories go confining. These vortices occur 
as dark energy (percolated by Kosterlitz-Thouless transitions) or as dark matter (depercolated vortex loops), the expansion of the universe converting the former 
to the latter. Today's dark energy $\Omega_{\Lambda}$ could be a homogeneous-field contribution 
of the PSA. Lacking for the time being a detailed ``microscopic'' understanding of the 
vortex-antivortex ensemble, we here assume for simplicity only one 
such instantaneous depercolation to occur at $z_p$. The addition of an emergent component of dark matter 
makes the dark sector more complex. However, as argued above, an interpolation between a smaller, early-time component of dark matter to the presently observed one is required as a direct consequence of the new $T$-$z$ 
relation of Eq.\,(\ref{TzT}). To posit a sudden transition from a dark-energy like component to the emergent dark-matter component is a simplification which may not survive the results of a detailed study of the physics of formely strongly bound vortices/antivortices in the associated percolate. It is clear though that if the mean separation between vortex/antivortex cores in the percolate is small compared to cosmological scales then the percolate does not exhibit any density contrast (like a cosmological constant). We imagine a vortex-antivortex-core interaction potential which is flat and positive for a certain range of distances but exhibits a barrier around a critical distance. To overcome this barrier energy must be released from the vortices'/antivortices' very, gravitating field configurations after which they move independently -- basically following the Hubble flow -- in the sense of a pressure-free fluid. According to our present fit results, see Table 2, this transition occurs even outside of reach of future, space based gravitational-wave observatories (LISA will cover redshifts up to $z=30$). Therefore, the here-introduced complexity of the dark-sector physics is undesirable in the sense of Occam's razor. We envisage though, that the (radiatively) predicted blackbody anomaly \cite{Hofmann2016a}, maximal for $T\sim 5\,$K, will be confirmed by terrestrial laboratory experiments. This would then strongly support the new $T$-$z$ relation (free quasi-particles) which enforces a modified dark sector similar to Eq.\,(\ref{edmdef}).

An important parameter of SU(2)$_\CMB$ is the fraction $f_p$ of primordial dark matter 
to today's dark matter   
\eqb
f_p\equiv\frac{\Omega_{\rm pdm,0}}{\Omega_{\rm pdm,0}+\Omega_{\rm edm,0}}\,.
\eqe 
Notice that in neglecting the baryonic contribution in the total matter density, Eq.\,(\ref{Omegacon}) implies $f_p \approx 1/4$.
In Fig.\,\ref{Fig-2} the redshift of matter-radiation equality $z_{\rm eq}$ is depicted as a function of 
$f_p$. In inspecting Eq.\,(\ref{zeqsu2}) this demonstrates a certain degree 
of matter domination at $z_{\SU,{\rm rec}}\sim 1700$ provided that $f_p$ is not too low.  

\subsubsection{The cosmological model SU(2)$_\CMB$\label{cosmomod}}

Now we are in a position to set up the entire cosmological model  SU(2)$_\CMB$ in terms of its Hubble parameter $H$
\begin{equation}\label{eq:def:hubbleCosmo}
H^2(z)  = H_0^2 \Big(\Omega_{\rm ds} (z) + \Omega_b (z) + \Omega_{{\YM}} (z) + \Omega_\nu (z) \Big)  \, ,
\end{equation}
where  $\Omega_{b}$ is the baryonic density parameter, $\Omega_{{\YM}}=\rho_\SU/\rho_c$ denotes the contribution of the SU(2) plasma, as defined in Eq.\,(\ref{eq:def:energyDensitySu2}) below, $\rho_c$ is the critical density $\rho_c \equiv 3 H_0^2/(8 \pi G)$, and $\Omega_{\nu}$ refers to the neutrino density parameter.
Since $\Omega_{b}$ is matter-like (cf. Eq.~\eqref{hzmatt}), its $z$-dependence reads
\begin{equation}\label{eq:def:omegab}
\Omega_{b} = \Omega_{b,0} \left(  z + 1 \right)^3 \, .
\end{equation} 
Here, $\Omega_{b,0}$ is today's baryonic density parameter which can be confronted with predictions from Big-Bang Nucleosynthesis (BBN) and with the results of baryon censuses. The contribution of the SU(2) plasma $\Omega_{{\YM}}$ comprises 
the massless mode ($\gamma$), identified with the U(1) photon, 
the massive quasi-particle modes ($V_\pm$), and the thermal ground state ($\text{gs}$).
The sum of their energy densities reads~\cite{Hofmann2016a}
\begin{equation}\label{eq:def:energyDensitySu2}
\rho_{\YM} =   \underbrace{2 \frac{T^4 }{2 \pi^2}   \tilde{\rho} (0) }_{\equiv \rho_{\YM ,\gamma}}+ \underbrace{6 \frac{T^4 }{2 \pi^2} \tilde{\rho} \left(2 M\right)}_{\equiv \rho_{\YM,V_\pm}} + \underbrace{\vphantom{\frac{T^4 }{2 \pi^2} } 4 \pi \Lambda_{\YM}^3 T}_{\equiv \rho_{\YM,\text{gs}}} \, ,
\end{equation}
where $\tilde{\rho}$ is given as 
\begin{equation}\label{eq:def:energyDensityBarDefinition}
\tilde{\rho} (y) = \int_0^\infty \mathrm d x \, x^2 \frac{\sqrt{x^2+y^2}}{\exp\left[-\sqrt{x^2+y^2}\right]-1} \,,
\end{equation}
the reduced mass $M$ reads
\begin{equation}\label{eq:def:reducedMassDef}
M = \frac{m_{V_\pm}}{2 T} \,,
\end{equation}
and $\Lambda_{\YM}$ denotes the Yang-Mills scale $\sim 10^{-4}$\,eV \cite{Hofmann2009}. 
To obtain the $V_\pm$ mass, we demand the Legendre transformation
\begin{equation}\label{eq:def:legendreTrafo}
\rho_{\YM} = T \frac{\mathrm d P_{\YM}}{\mathrm d T} - P_{\YM}\,,
\end{equation} 
implying 
\begin{equation}\label{cond:def:legendreCond}
0 \overset{!}{=} \frac{\partial P_\YM}{\partial m_{V_\pm}} \,.
\end{equation} 
Here the pressure $P_{\YM}$ is defined as 
\begin{equation}\label{eq:def:pressureSu2}
P_{\YM}(\lambda,M) = -\Lambda_{\YM}^4 \left(\frac{2 \lambda^4}{(2\pi)^6} \left[2 \tilde{P}(0)+6 \tilde{P}(2M) \right] + 2 \lambda \right) \,,
\end{equation}
with the dimensionless integral
\begin{equation}
\tilde{P}(y) = \int_0^\infty \mathrm d x \, x^2\log \left( 1 - \exp \left[- \sqrt{x^2 + y^2} \right] \right) \,,
\end{equation}
and the dimensionless temperature
\begin{equation}
\lambda = \frac{2 \pi T}{\Lambda_{\YM}} \,.
\end{equation}
Eq.\,\eqref{cond:def:legendreCond} was solved for $m_{V_\pm}(T)$ in \cite{Hofmann2016a}.
Note that $\Omega_{\YM, \gamma}(T)= \Omega_{\LC,\gamma}(T)$, but that $\Omega_{\YM, \gamma}(z) \neq \Omega_{\LC,\gamma}(z)$ because of Eq.\,\eqref{TzT}.
Finally, we have
\begin{equation}\label{eq:def:omegaNu}
\Omega_\nu(z) = \frac{7}{8} N_{\rm eff}\,\left(\frac{16}{23} \right)^{\frac{4}{3}} \Omega_{\YM, \gamma}  (z) \, ,
\end{equation}
where a modified factor for the conversion of neutrino to CMB temperature occurs 
because of the additional relativistic degrees of freedom during $e^+e^-$ annihilation~\cite{Hofmann2015}.

\subsubsection{Sound velocity $c_s$ and contributions to $R$}

\begin{figure}
\centering
\includegraphics[width=\linewidth]{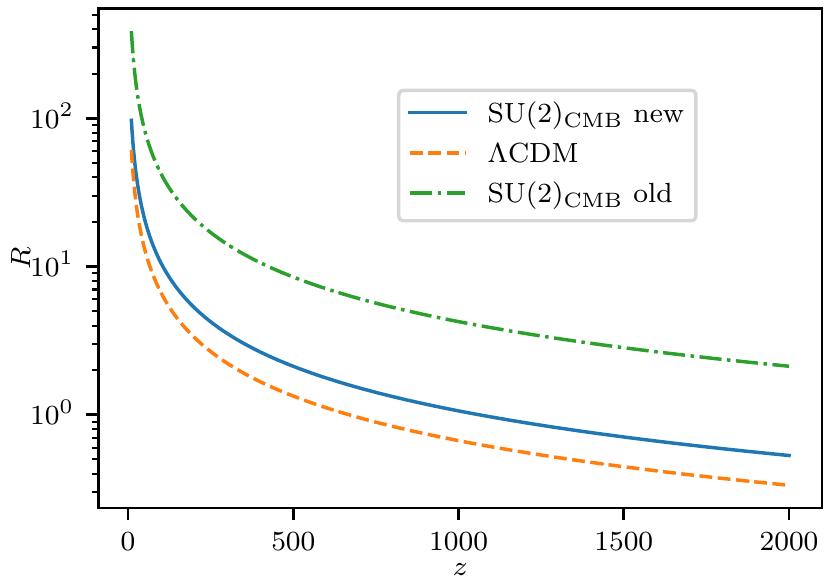}
\caption{$R$ as a function of $z$ as erroneously defined in \protect\cite{HH2017} 
(${\rm SU(2)}_\CMB$ old), as correctly used in the present work 
(${\rm SU(2)}_\CMB$ new), and as utilised in $\Lambda$CDM.
\protect{\label{Fig-3}}}      
\end{figure}

The co-moving sound horizon $r_s(z)$, which is a high-$z$ variable ($z\ge z_\text{rec}$), is given as a 
functional of the Hubble parameter $H(z)$ and the sound velocity $c_s(z)$ 
\eqb
\label{rsz}
r_s(z)\equiv\int_z^\infty dz^\prime\,\frac{c_s(z^\prime)}{H(z^\prime)}\,,
\eqe 
where $c_s$ is represented by
\eqb
\label{soundcs}
c_s(z)\equiv\frac{1}{\sqrt{3(1+R(z))}}\,.
\eqe 
In $\Lambda$CDM the ratio $R_\LC$ can be expressed in terms of 
entropy densities $s_\LC$ or re-scaled energy densities $\rho_\LC$ of baryons (b) and photons ($\gamma$), 
\eqb
\label{defRLam}
R_{\LC}\equiv\frac{s_{b}(z)}{s_{\LC,\gamma}(z)}=\frac34\frac{\rho_{b}(z)}{\rho_{\LC,\gamma}(z)}\quad (z \gg 1)\,.
\eqe  
Since the entropy or energy densities in the definition of $R$ relate to conserved fluids the generalisation of Eq.\,(\ref{defRLam}) to the baryon-Yang-Mills 
plasma replaces $s_{\LC,\gamma}(z)$ or $\rho_{\LC,\gamma}(z)$ by  $s_{\YM}(z)$ or $\rho_{\YM}(z)$, 
respectively, to define $R_{\YM}$. In \cite{HH2017} this was overlooked. In Fig.\,\ref{Fig-3} $R$ is plotted as a function of $z$ 
for the definition erroneously used in \cite{HH2017}, the one of the present work, which considers the full Yang-Mills plasma, and the one of $\Lambda$CDM.    

\subsection{First order in cosmological perturbations\label{FOCP}}

Eq.\,(\ref{eq:def:hubbleCosmo}) represents the homogeneous and isotropic 
cosmological background. Based on this definition, we now briefly discuss 
the evolution of linear cosmological perturbations as induced 
by primordial, scalar, and adiabatic curvature perturbations. In conformal Newtonian 
gauge \cite{MB1995} we consider metric perturbations subject to the 
line element
\eqb
\label{lineelemnt}
\mathrm d s^2 = a(\eta)^2 \left( -(1+2\psi) (\mathrm d \eta)^2 + (1- 2 \phi) \mathrm d x^i \mathrm d x^i \right) \, ,
\eqe  
where $\eta$ denotes the conformal time, defined by
\begin{equation}
\label{eq:def:conformaltime}
\mathrm d \eta = \frac{\mathrm d t}{a} \, ,
\end{equation}
and $t$ is the physical time.
We only consider scalar metric perturbations in terms of $\psi$ and $\phi$.
The derivation of the evolution equations for $\psi$ and $\phi$ is standard~\cite{MB1995}.
Perturbations $\Psi$ in the phase-space distribution-functions $f$ for our relativistic degrees of freedom ($N_\text{eff}$ species of neutrinos\footnote{In the present version of SU(2)$_\CMB$, we refrain from considering neutrino masses.}, photons, and $V_\pm$) are governed by the first-order terms of their respective Boltzmann equations.
However, we require a modified dispersion law\footnote{Considering a 
modification of $p_\mu p^\mu$ in the Boltzmann equations for photons and $V_\pm$ due to radiative 
effects, plays out at low $z$ (and therefore low $l$) only, and is neglected in our present treatment.}
\begin{equation}
\label{eq:def:modDispersion}
p_\mu p^\mu = m_{V_\pm}^2 \,
\end{equation} 
where $p_\mu$ is the physical four-momentum, and $m_{V_\pm}$ denotes $V_\pm$'s quasi-particle mass.
As a consequence of Eq.\,\eqref{eq:def:modDispersion}, $V_\pm$'s zeroth-order Bose-Einstein distribution function $f_0$ is modified compared to the massless case.

\subsubsection{Initial conditions for emergent dark matter}

In evolving emergent dark matter, modelled by an instantaneous 
depercolation of Planck-scale axion vortices at $z_p$, as described by Eq.\,\eqref{edmdef}, the according evolution equations in co-moving $k$ space,
\eab
\label{dmevol}
\dot{\delta}_{\rm edm}&=&-\theta_{\rm edm}+3\dot{\phi}\,,\nonumber\\ 
\dot{\theta}_{\rm edm}&=&-\frac{\dot{a}}{a}\theta_{\rm edm}+k^2\psi\,,
\eae
need to be supplemented with initial conditions for the density contrast ${\delta}_{\rm edm}=\frac{\delta\rho_{\rm edm}}{\rho_{\rm edm}}$ and the 
divergence $\theta_{\rm edm}\equiv ik^jv_{\text{edm},j}$ of the fluid velocity $\vec{v}_\text{edm}$ at conformal time $\eta_p$ which  
associates with the depercolation redshift $z_p$. Dots refer to differentiation w.r.t. $\eta$.
In the computation of the angular power spectra we observer that strong variations,
\eqb
\label{inidelta}
{\delta}_{\rm edm}(\eta_p)=\alpha\,{\delta}_{\rm pdm}(\eta_p)\quad (0\le\alpha\le 1)\,,
\eqe
do not influence the result. On the other hand, variations in the 
initial condition for ${\theta}_{\rm edm}$, 
\eqb
\label{initheta}
{\theta}_{\rm edm}(\eta_p)=\beta\,{\theta}_{\rm pdm}(\eta_p)\quad (0\le\beta\le 1)\,
\eqe
influence the power at low $l$, $l\le 30$. This conclusion 
is drawn from setting $\alpha=\beta=1$, minimising $\chi^2$ with respect to Planck's TT, TE, and EE 
power spectra, and subsequently varying $\alpha$ and $\beta$ with the other cosmological parameters kept fixed. 
In actual fits we therefore set $\alpha=1$ and consider $\beta$ to be a free parameter. 

The important feature of depercolated configurations of the Planck-scale axion field is that their individual 
extent (both transverse and along the vortex loop) is small on 
cosmological scales. Note that this may not be the case on astrophysical scales. Vortex loops (and other texture in association with the Planck-scale axion field) thus can be 
considered non-relativistic particles in isolation with a cosmological equation of state $P=0$. 
Since our simple transition of Eq.\,(\ref{edmdef}) from a dark-energy like component (a percolate of these field configuration) into an  ideal gas of them in isolation so far only represents a parametrisation of our ignorance concerning the ``microscopics'' of the percolated state, including 
the field configurations in individual self-gravitating vortex cores, there is not yet a handle on usefully discussing causality in the initial conditions of the fluctuations in energy density and the divergence of peculiar fluid velocity in the emergent dark matter. Our initial conditions (\ref{inidelta}) and (\ref{initheta}) assume that the according modes follow, up to a rescaling of order unity, the one of the primordial dark matter, evolved from Gaussian initial conditions throughout the linear regime. (According to Tab.\,\ref{tab:CosmParamFinal} and Fig.\,\ref{H0scatter} it is justified to assume that $z_p>50$). Since these exhibit zero means the fluctuations of the emergent dark matter also exhibit zero means, and energy conservation therefore is guaranteed.

\subsubsection{Boltzmann hierarchy for $V_\pm$}\label{sec:boltzmannstuff}

We now discuss a number of points to be considered when deriving the Boltzmann 
hierarchy for $V_\pm$. Because we neglect radiative corrections in 
deconfining SU(2) Yang-Mills thermodynamics only the collisionless Boltzmann equation for the phase-space distribution $f$ is relevant. It reads    
\begin{equation}
\label{genBE}
\frac{\mathrm d f}{\mathrm d \eta} = \frac{\partial f}{\partial \eta} + \frac{\mathrm d x^i}{\mathrm d \eta} \frac{\partial f}{\partial x^i} + \frac{\mathrm d q}{\mathrm d \eta} \frac{\partial f}{\partial q} + \frac{\mathrm d n_i}{\mathrm d \eta} \frac{\partial f}{\partial n_i} = 0\,,
\end{equation}
where the co-moving momentum $\vec{q}$ and its modulus $q$ define $\vec{n}\equiv\frac{\vec{q}}{q}$.

The only non-standard term involves the factor $\frac{\mathrm d q}{\mathrm d \eta}$. As usual, 
the perturbation $\Psi$ of the unperturbed phase-space distribution $f_0=1/(\exp(\epsilon/T_0)-1)$ is introduced as
\begin{equation}
f = f_0(\epsilon) \left(1 + \Psi\right)\,,
\end{equation}
where the co-moving energy $\epsilon$ reads
\begin{equation}
\label{comovenem>0}
\epsilon \equiv \left(q^2 + a^2 \frac{m_{V_\pm}^2}{{\cal S}^2\, T_0^2}\right)^{1/2}\,.
\end{equation}
The scaling function $\mathcal{S}$ is defined in
Eq.\,(\ref{devlinS}). Since the term $\frac{\mathrm d q}{\mathrm d \eta}$ is determined by the geodesic 
equation for a massive point particle, we may write
\begin{equation}
\begin{split}
\frac{\mathrm d q}{\mathrm d \eta} \frac{\partial f}{\partial q} = &\left(q \dot{\phi} - \epsilon n_i \partial_i \psi - \frac{a^2 m_{V_\pm} \dot{m}_{V_\pm}}{q} \right) \times \\ 
&\left( \frac{\partial f_0}{\partial q} (1 + \Psi) + f_0 \frac{\partial \Psi}{\partial q}\right)\,.
\end{split}
\end{equation}
The use of the geodesic equation for a quasi-particle must be questioned, if this particle associates with pure quantum fluctuations \cite{Hofmann2017}. If at all, temperature fluctuations in the $V_\pm$ sector can 
thus only be coherently propagated via the low-frequency regime of $\gamma$ fluctuations in terms of classical electromagnetic waves \cite{Hofmann2016b}. To do this, a residual interaction between $V_\pm$ and $\gamma$ is 
required. Albeit such an interaction occurs \cite{Hofmann2016a}, its efficiency in conveying the coherent 
propagation of $V_\pm$ temperature fluctuations must be questioned, especially at high temperatures \cite{Falquez2010}. 
To ignore the $V_\pm$ Boltzmann equations and associated source terms in linearised Einstein equations 
thus is a physically motivated option. On the other hand, considering the evolution of $V_\pm$ temperature 
fluctuations via the coherently propagating low-frequency sector in $\gamma$ implies 
that $\epsilon=q$ in the $V_\pm$ geodesic equation. At the same time, $m_{V_\pm}>0$ is required in $f_0$. 
Since the structure of temperature fluctuations is mainly imprinted before and during recombination, setting $m_{V_\pm}=0$ in the geodesic 
equation does not influence the prediction for the power spectra in practice. 
Note that due to Eq.\,(\ref{comovenem>0}) an explicit dependence of $f_0$ on $\eta$ needs to be 
considered via $a=a(\eta)$. Transforming Eq.\,(\ref{genBE}) into $k$ space and otherwise following 
the standard procedure of linear perturbation theory \cite{MB1995}, one arrives at the following hierarchy    
\begin{align}
\dot{\Psi}_0 &= - k\Psi_1 - \frac{\mathrm d \ln f_0}{\mathrm d \ln q}\dot{\phi}  - \frac{1+\Psi_0}{f_0} \frac{\partial f_0}{\partial \eta}\,, \\
\dot{\Psi}_1 &= \frac{k}{3 }(\Psi_0 - 2 \Psi_2) - \frac{1}{3} \frac{\mathrm d \ln f_0}{\mathrm d \ln q} k \psi - \frac{\Psi_1}{f_0} \frac{\partial f_0}{\partial \eta}\,, \\
\dot{\Psi_l} &= \frac{k}{(2l+1)}\left[l \Psi_{l-1}-(l+1)\Psi_{l+1} \right] -\frac{\Psi_l}{f_0} \frac{\partial f_0}{\partial \eta}\,, 
\end{align}
where the $\Psi_l(\vec{k},q,\eta)$ are the expansion coefficients for $\Psi(\vec{k},\hat{n},q,\eta)$ into Legendre polynomials.

Since the $V_\pm$ are free quasi-particles in the here-considered approximation, their initial conditions are implemented in analogy to those of massive neutrinos \cite{MB1995}. 

\section{Constraints from standard ruler and $\boldsymbol{\theta_*}$}\label{SHBDtheta}

The angular size $\theta_*$ of the sound horizon $r_s$, see Eq.\,(\ref{rsz}), at photon decoupling $z_*$ is defined as 
\begin{equation}\label{eq:def:angularSize}
\theta_* \equiv \theta (z_*) \equiv \frac{1}{z_*+1} \frac{r_s(z_*)}{d_A(z_*)} \,,
\end{equation}
where the angular diameter distance $d_A(z)$ reads
\begin{equation}
d_A (z) \equiv \frac{1}{z+1} \int_0^z \mathrm d z' \, \frac{\mathrm d z'}{H(z')} \,.
\end{equation}
Therefore, $\theta_*$ is an all-$z$ variable.
Conventionally, $z_*$ denotes the redshift at which the optical depth $\tau_*$ is unity: 
\begin{equation}\label{cond:def:methodOneStar}
1 \overset{!}{=} \tau_*(z_*) \equiv \int_0^{z_*}  \frac{\mathrm d z}{H(z)} \, \dot\tau_* (z) \,.
\end{equation} 
Here, $\dot \tau_*(z)$ is given as
\begin{equation}\label{eq:def:opticalDepthDerivation}
\dot\tau_* (z) \equiv \sigma_T \frac{n_e(z)}{z+1} \,,
\end{equation}
where $\sigma_T$ denotes the Thomson cross section, and $n_e$ is the number density of free electrons.
Analogously, the redshift of baryon drag $z_d$ is conventionally obtained by replacing $\dot\tau_* (z)$ with \cite{Hu1994,Hu1996} 
\begin{equation}\label{cond:def:baryDragDepthDerivation}
\dot\tau_d (z) \equiv \frac{1}{R(z)} \dot\tau_* (z) 
\end{equation}
in Eq.~\eqref{cond:def:methodOneStar}, where $R(z)$ is specific to the cosmological model under investigation, compare with Eq.\,\eqref{defRLam} for example.
In~\cite{HH2017} it was argued that due to the finite peak width in $z'$ of the visibility function $D_* (z,z')$ photon freeze-out occurs at its \emph{left flank}. Function $D_* (z,z')$ is given as
\begin{equation}
D_*(z,z') \equiv \frac{\dot\tau_* (z')}{H(z') (z'+1)} \exp \left(-\left[ \tau_*(z) - \tau_*(z') \right] \right) \,.
\end{equation}
To determine $z_d$, an analogous argument relates to decoupling at the \emph{left flank} of $D_d(z,z')$.
Notice that, with the conventional decoupling conditions, the values of $z_*$ and $z_d$ practically coincide with the 
peak positions of their visibility functions $D_*(z,z')$ and $D_d(z,z')$.

\begin{figure*}
\centering
\includegraphics[width=\textwidth]{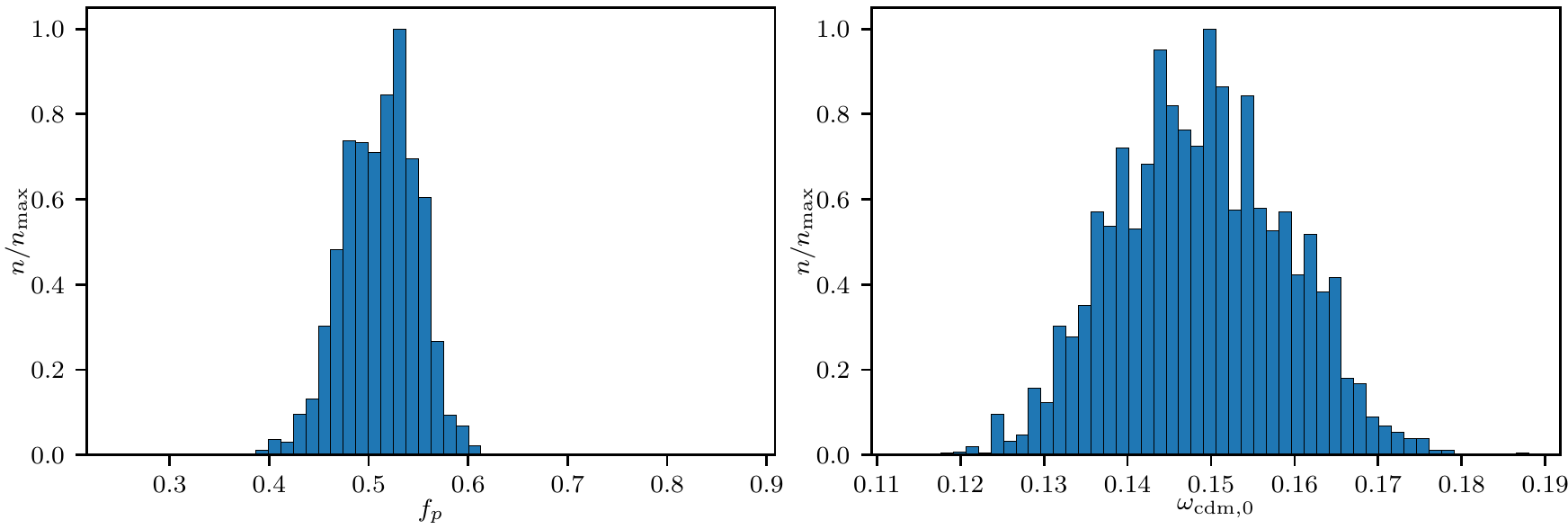}
\caption{\protect{\label{peaks}}
Normalised histograms for $f_p$ (left panel) and $\omega_{\rm cdm,0}$
(right panel) generated from the \emph{MC-MH} algorithm. 
We use 500000 steps and $H_0=73.48\,$\,km\,s$^{-1}$Mpc$^{-1}$ \protect\cite{Riess2018}.
}
\end{figure*}

To compute $r_d\equiv r_s(z_d)$ and $\theta_*$ in SU(2)$_\CMB$, six parameters need to be prescribed.
These are the helium fraction in baryon mass $Y_p$, the scaled baryon-to-photon number fraction $\eta_{10}$, $N_\text{eff}$, $z_p$, $f_p$, and the physical energy density
\eqb
\label{omtot}
\omega_{\rm cdm,0}\equiv h^2 \left(\Omega_{\rm edm,0}+\Omega_{\rm pdm,0}\right)\,,
\eqe
where $H_0\equiv h\cdot 100$\,km\,s$^{-1}$Mpc$^{-1}$. 
While $\theta_*$ is nearly model-independently and accurately determined by the observed peak 
structure in the TT angular power spectrum \cite{Boomerang,wmap2003,Ade2016}, 
\eqb
\label{exptheta*}
100\cdot\theta_*=1.04122\pm 0.00045\,,
\eqe
there is no such consensus about the value of the 
standard ruler $r_d$. The value of $r_d$ associates with a bumpy feature in the matter correlation function, introduced by 
baryonic acoustic oscillation (BAO) \cite{Heavensetal2014,Bernal2016}, and is connected to $H_0$ via $h r_d=\mbox{const}\sim 100$\,Mpc \cite{Heavensetal2014,Bernal2016}. Our strategy in exploring the implications of 
SU(2)$_\CMB$ is to prescribe value for $\theta_*$ as in Eq.\,(\ref{exptheta*}) as well as $r_d$ by minimising the cost function
\begin{equation}\label{eq:def:costFunctionForMc}
C (r_d,\theta_* ; r_{d,o}, \theta_{*,o}) = \frac{\left( \theta_* - \theta_{*,o} \right)^2}{2 \sigma^2_{*,o}} + \frac{\left( r_d - r_{d,o} \right)^2}{2 \sigma^2_{d,o}} \,.
\end{equation}
Here, $r_{d,o}$, and $\theta_{*,o}$ refer to the prescribed values, and $\sigma_{d,o}$, and $\sigma_{*,o}$ are typical errors in their observational extractions.
We use $\sigma_{*,o} = 0.00045/100$ (see Eq.~\eqref{exptheta*}) and $\sigma_{d,o}$ as inferred from a 
propagated error in $H_0$ of $\Delta {H_0} \sim 1$ km\,s$^{-1}$Mpc$^{-1}$~\cite{Bernal2016}. Moreover, $\theta_{*,o}$ is always fixed 
to the value given in Eq.\,(\ref{exptheta*}) while  
$r_{d,o}$ is varied in the range $135\,$Mpc$\,\le r_{d,o}\le 147\,$Mpc~\cite{Bernal2016}. 
In performing the analysis, we use a Monte Carlo (MC) procedure which is inspired by the Metropolis-Hastings  algorithm~\cite{Metropolis1953,Hastings1970}, dubbed here \emph{MC-MH}.
For all of the above listed parameters of SU(2)$_\CMB$ we consider the following prior ranges
\begin{equation}\label{priorrange}
\begin{split}
0 &\leq f_p \leq 1\,, \quad &100 &\leq z_p \leq 900 \,, \\ 
5.8823 &\leq \eta_{10} \leq 6.6823 \,,\quad &0.2 &\leq Y_P \leq 0.3 \,, \\
3.0 &\leq N_\text{eff} \leq 3.2 \,, \quad &0.1126 &\leq \omega_{\rm cdm,0} \leq 0.1886 \,.
\end{split}
\end{equation}
The large ranges of $f_p$ and $z_p$ are motivated by our present ignorance about the ``microscopics'' of Planck-scale-vortex depercolation, those of $Y_P$ and $\eta_{10}$ contain the BBN bounds, $N_\text{eff}$ is tightly constrained by particle physics on flavour neutrinos ~\cite{Tanabashi2018} and neglecting sterile neutrinos. The prior range for $\omega_{\rm cdm,0}$ is bounded from below by the $\Lambda$CDM fit to the 2015 Planck data, but we allow for a generous increase as a result of our present analysis.
In step $N$ and assuming uniform distributions, we compute a point in the six-dimensional parameter space spanned by ~\eqref{priorrange}, calculate its values of $r_d$ and $\theta_*$, and appeal to the following decision function
\begin{equation}\label{eq:def:decisionFunctionForMc}
D (r_d,\theta_* ; r_{d,o}, \theta_{*,o}) = \exp \left[ - C(r_d,\theta_* ; r_{d,o}, \theta_{*,o}) \right] \, ,
\end{equation}
storing this point when the condition
\begin{equation}
 \frac{D_{N}}{D_{N-1} } > U (0,1) 
\end{equation}
is satisfied.
Here, $U(0,1)$ is a randomly generated value between zero and unity.
For each parameter a histogram is thus generated from the so-obtained set of points.
The histogram is composed of fifty bins within each parameter range.
If a peak structure emerges, a Gaussian, whose peak position is interpreted as the preferred value, is fitted.
The width of this Gaussian serves as an estimate of the error.
Note that such an estimate depends on the prior ranges.
Single peaks only occur for $f_p$ and $\omega_{\rm cdm,0}$ (see Fig.~\ref{peaks}). 
For all other parameters the histograms are nearly flat.
Therefore, the model determines the values of two out of its six parameters in terms of the two independent quantities $r_d$ and $\theta_*$.
Our analysis may thus be restricted to the investigation of the dependencies of $f_p$ and $\omega_{\rm cdm,0}$ on $r_{d,o}$. 

\begin{figure*}
\centering
\includegraphics[width=\textwidth]{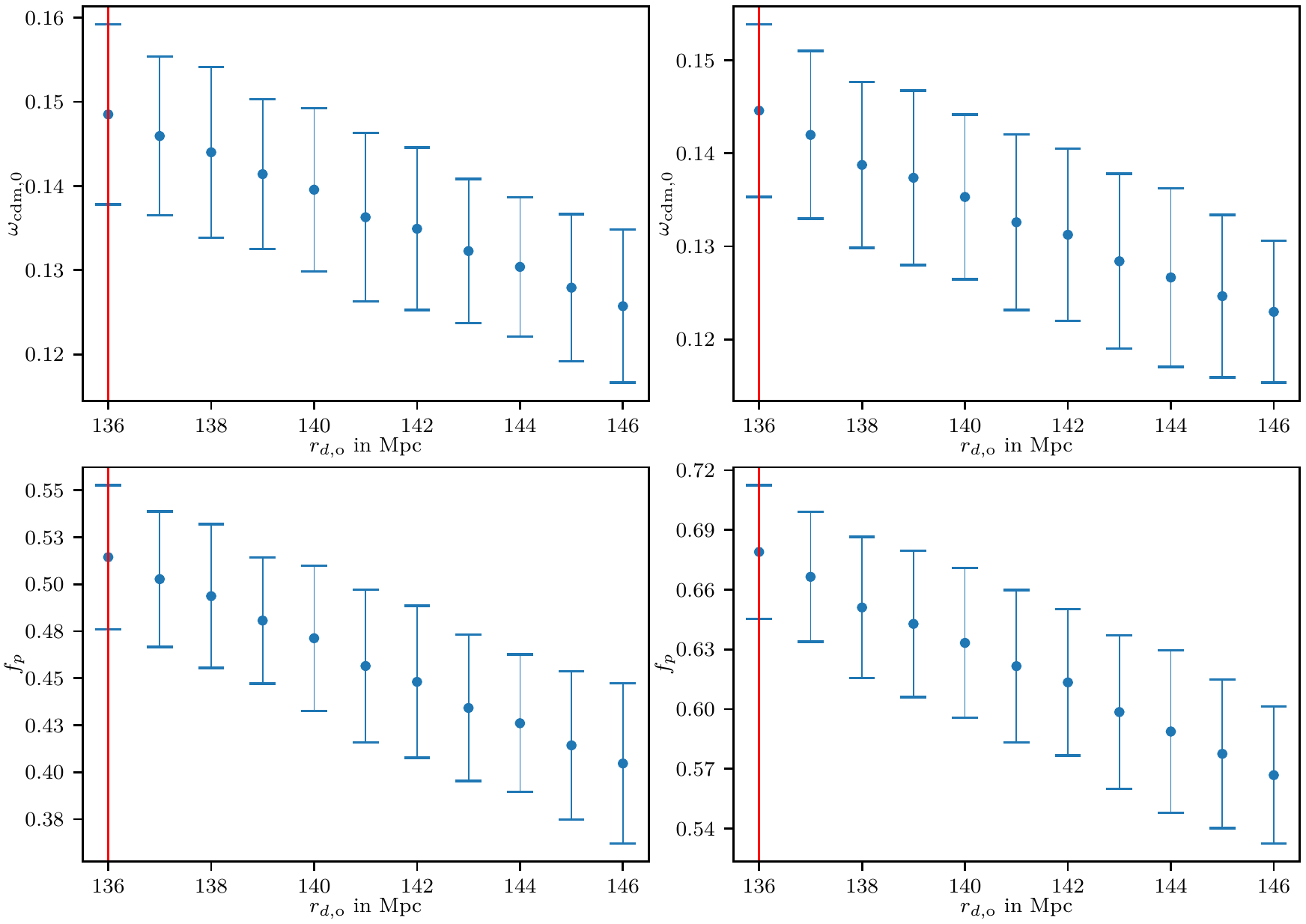}
\caption{\protect{\label{affinefigs}}
Preferred values of $\omega_{\rm cdm, 0}$ (upper panels) and $f_p$
(lower panels) depending on the value of $r_{d,o}$ generated from the
\emph{MC-MH} algorithm for the conventional (left panels) and \emph{left flank} (right panels) extraction method.
In each case, we find affine dependencies.
The vertical red line depicts the value of $r_{d,o}$ corresponding to
$H_0 = 73.48\,$\,km\,s$^{-1}$Mpc$^{-1}$ ~\protect\cite{Riess2018}.
}
\end{figure*}

\begin{figure*}
\centering
\includegraphics[width=\textwidth]{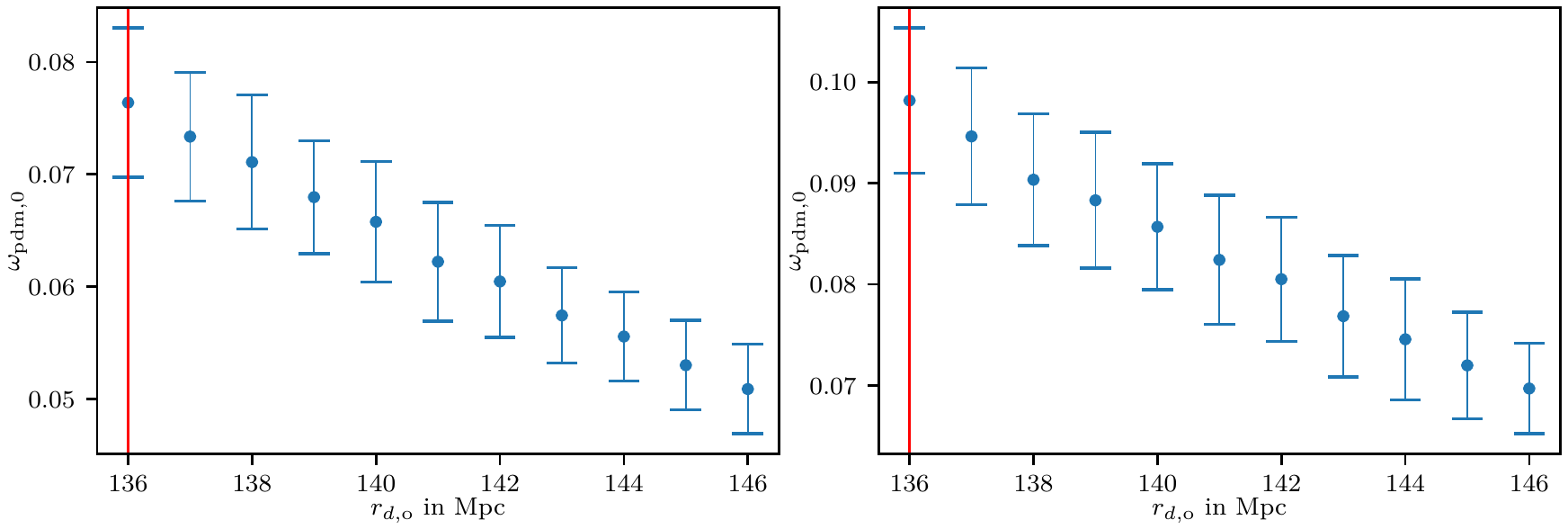}
\caption{\protect{\label{omegapdm}}
Preferred values of $\omega_{\rm pdm, 0}=\omega_{\rm cdm,0} f_p$ for
the conventional (left panel) and \emph{left flank} (right panel)
extraction method, errors propagated from those of
$\omega_{\rm cdm,0}$ and $f_p$.
}
\end{figure*}

We find approximate affine dependencies on $r_d$ for both the conventional and \emph{left flank} method discussed above, see Fig.~\ref{affinefigs}.
Notice that this predicts a curved function $\omega_{\rm pdm,0}=\omega_{\rm cdm,0} f_p$, see Fig.\,\ref{omegapdm}. 
At $r_{d,o}=136$\,Mpc, which corresponds to the typical local value $H_0 = (73.48\pm 1.66)$\,km\,s$^{-1}$Mpc$^{-1}$ \cite{Riess2018}, we obtain peak positions $\omega_{\rm pdm, 0}=0.077$ (conventional) and $\omega_{\rm pdm, 0}=0.098$ (\emph{left flank}).  
From Fig.~\ref{affinefigs} we extract peak positions $\omega_{\rm cdm, 0}=0.148$ (conventional) and $\omega_{\rm cdm, 0}=0.144$ (\emph{left flank}) at the same value of $r_{d,o}$. 
Compared to $\omega_{\rm cdm, 0}=0.1186\pm0.0020$ from the Planck best-fit (TT+lowP+lensing) to $\Lambda$CDM \cite{Ade2016} 
this is high. 

\section{Constraints from CMB angular power spectra\label{FO}}

Instead of confronting SU(2)$_\CMB$ with only two observationally imposed constraints, we now determine all of its parameters by means of the angular power spectra TT, TE, and EE \cite{Ade2016}.
To this end, we worked with the Cosmic Linear Anisotropy Solving System (CLASS) \\
\cite{CLASS} because 
its modular and well-documented structure allows for a targeted implementation of the modifications demanded by SU(2)$_\CMB$.

\subsection{Code adjustments in CLASS} 

Modifications are implemented in the following modules\footnote{The modified code can be obtained from the authors upon request.}: 
\texttt{input}, \texttt{background}, \texttt{thermodynamics}, \texttt{perturbation}, \texttt{output}. 
Also, we have introduced an additional module \texttt{unconventional}. 

In detail, \texttt{input} now contains the definitions of the additional cosmological parameters as introduced in Sec.\,\ref{cosmomod}: $z_p,\,\omega_{\rm edm,0},\, \alpha,\, \beta$, and the modified conversion between neutrino temperature $T_\nu$ and CMB temperature $T$. 
The thermodynamic quantities of the SU(2) Yang-Mills plasma, see Eqs.\,(\ref{eq:def:energyDensitySu2}) and (\ref{eq:def:pressureSu2}), and function ${\cal S}(z)$ of Eq.\,(\ref{devlinS}), are computed in \texttt{unconventional}. 
In \texttt{background} the cosmological model is defined according to Eq.\,(\ref{eq:def:hubbleCosmo}). 
We also introduce the unperturbed Bose-Einstein distribution for $V_\pm$ and $R_\YM$, the latter serving as an input in various other modules, e.g., in \texttt{thermodynamics} to define photon and baryon decoupling or in \texttt{perturbation} to define the photon-baryon scattering term in the Boltzmann equation. 
In \texttt{thermodynamics} we implement the modified $T$-$z$ relation in terms of function ${\cal S}(z)$. 
The Boltzmann hierarchy for $V_\pm$ and the Euler equation for emergent dark matter (\ref{dmevol}), including the model for the initial conditions of Eqs.\,(\ref{inidelta}) and (\ref{initheta}), are now added to \texttt{perturbation}. 
The module \texttt{output} is extended to accommodate the additional parameters of SU(2)$_\CMB$.

\subsection{Best-fit results: Planck data vs. SU(2)$_\CMBbold$ and $\Lambda$CDM}

To estimate effects of including/excluding 
the $V_\pm$ in the perturbation equations, 
see Sec.~\ref{sec:boltzmannstuff}, we consider the following two variants of the cosmological model.\\
\vspace*{0.2cm}
\noindent\underline{\textbf{$\text{SU(2)}_\CMBbold$ (physically favoured):}} \\ background model of Eq.\,(\ref{eq:def:hubbleCosmo}), disregarding coherent temperature fluctuations as induced by the $V_\pm$; account for $R_{\YM}$ in all other instances.\\
\vspace*{0.2cm}
\underline{\textbf{$\text{SU(2)}_\CMBbold$+$V_\pm$ (physically disfavoured):}} \\ background model of Eq.\,(\ref{eq:def:hubbleCosmo}) together with participation of coherent temperature fluctuations as induced by the $V_\pm$ in terms of their own Boltzmann hierarchy and associated source terms in evolution of metric perturbations; account for $R_{\YM}$ in all other instances.
\vspace*{0.2cm}

\begin{figure*} 
\begin{subfigure}{0.49\textwidth}
\includegraphics[width=\textwidth]{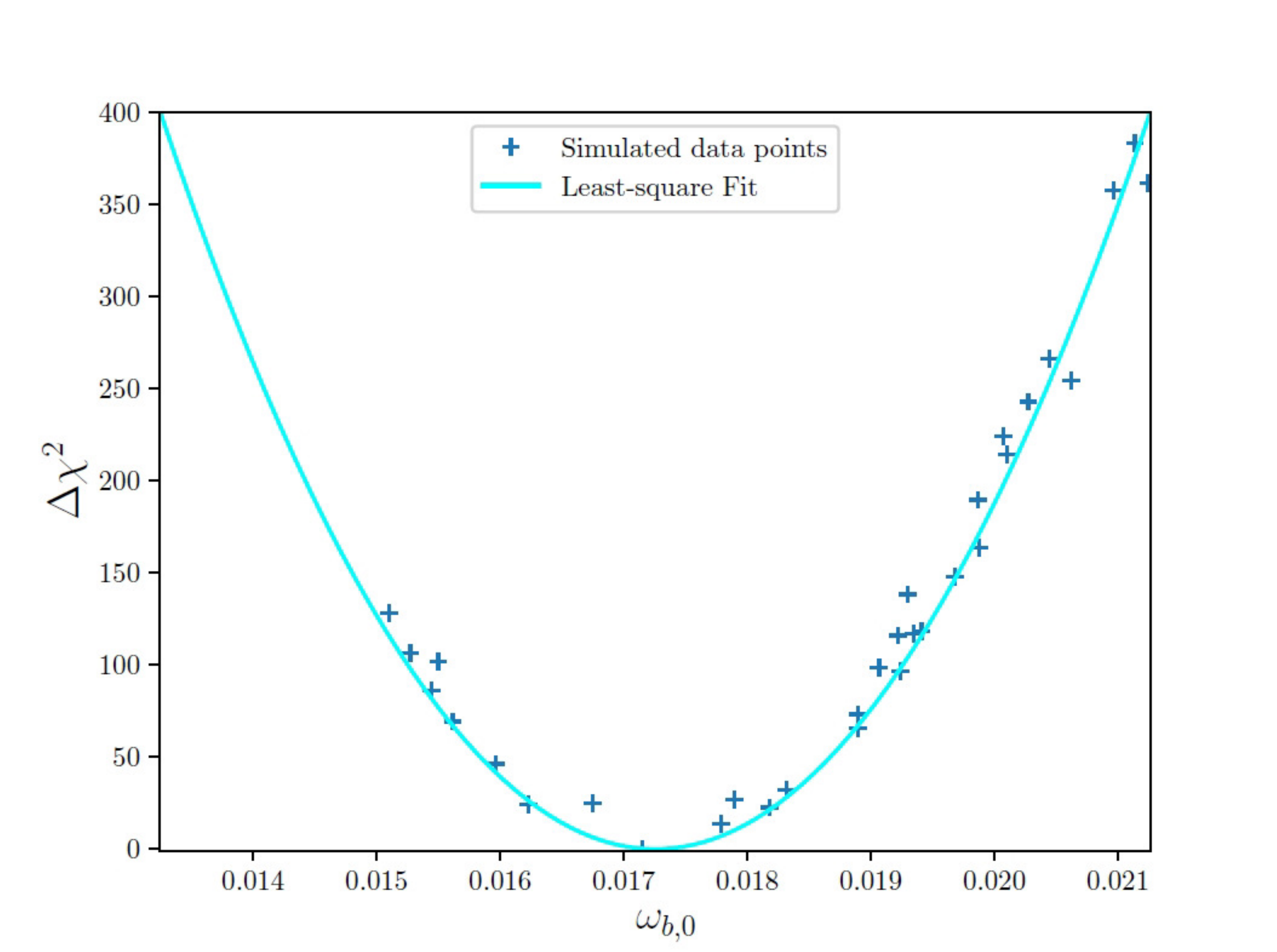}
\end{subfigure}\hspace*{\fill}
\begin{subfigure}{0.49\textwidth}
\includegraphics[width=\textwidth]{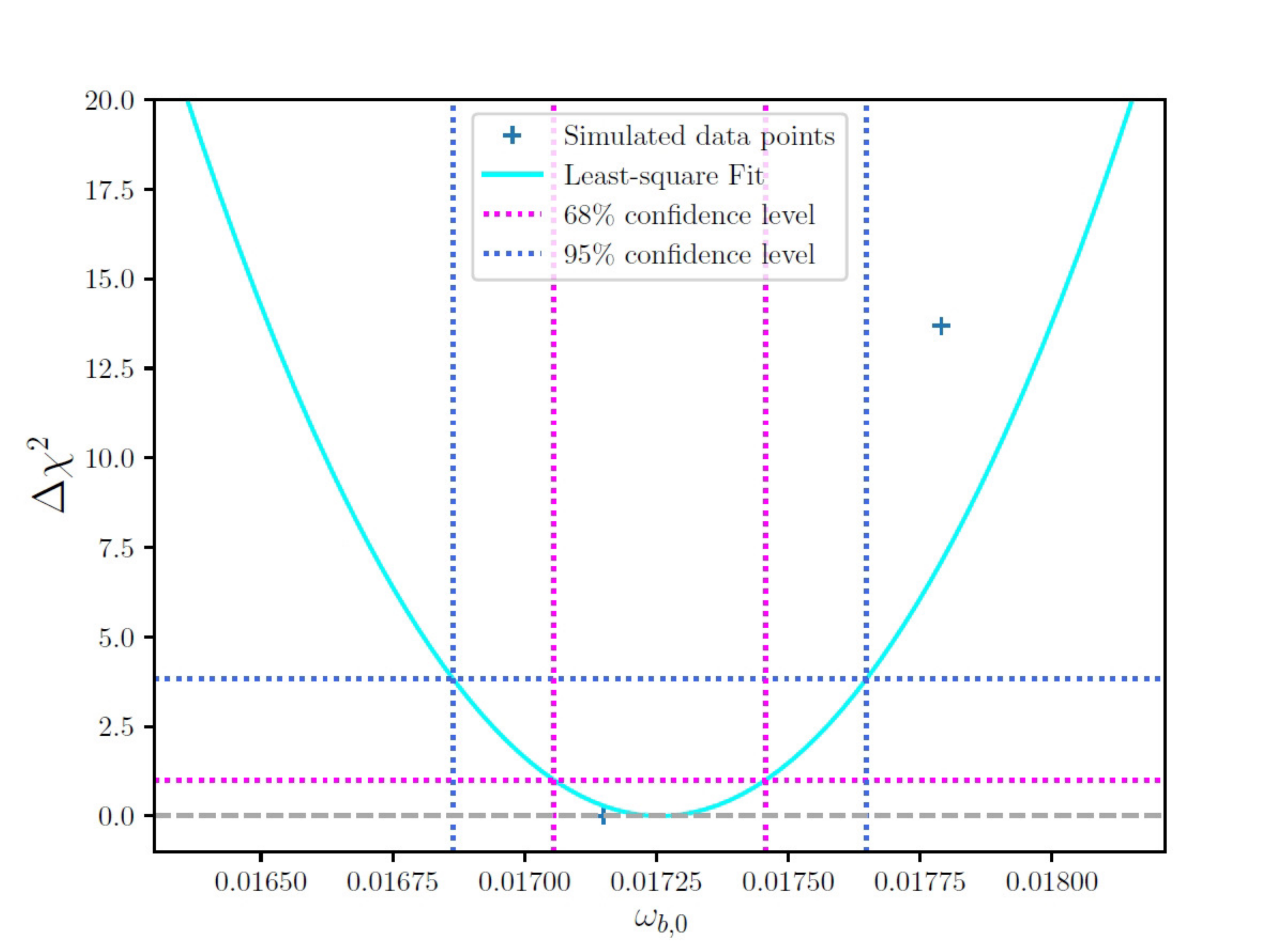}
\end{subfigure}
\caption{\protect{\label{fig_67_1}}Chi-square co-profile w.r.t. 
HiLLiPOP+lowTEB+lensing likelihood of the parameter $\omega_\text{b,0}$ together with the least-square fit 
to a parabola for $\text{SU(2)}_\CMB$ (left panel). Zoomed-in co-profile with intersections at $\Delta \chi^2 = 1, 3.84$ corresponding to 68\% and 95\% confidence levels (right panel). Central value and error range (68\% confidence) are quoted in the first column of Tab.\,\ref{tab:CosmParamFinal}.}

\end{figure*}

\begin{figure*}
\begin{subfigure}{0.49\textwidth}
\includegraphics[width=\textwidth]{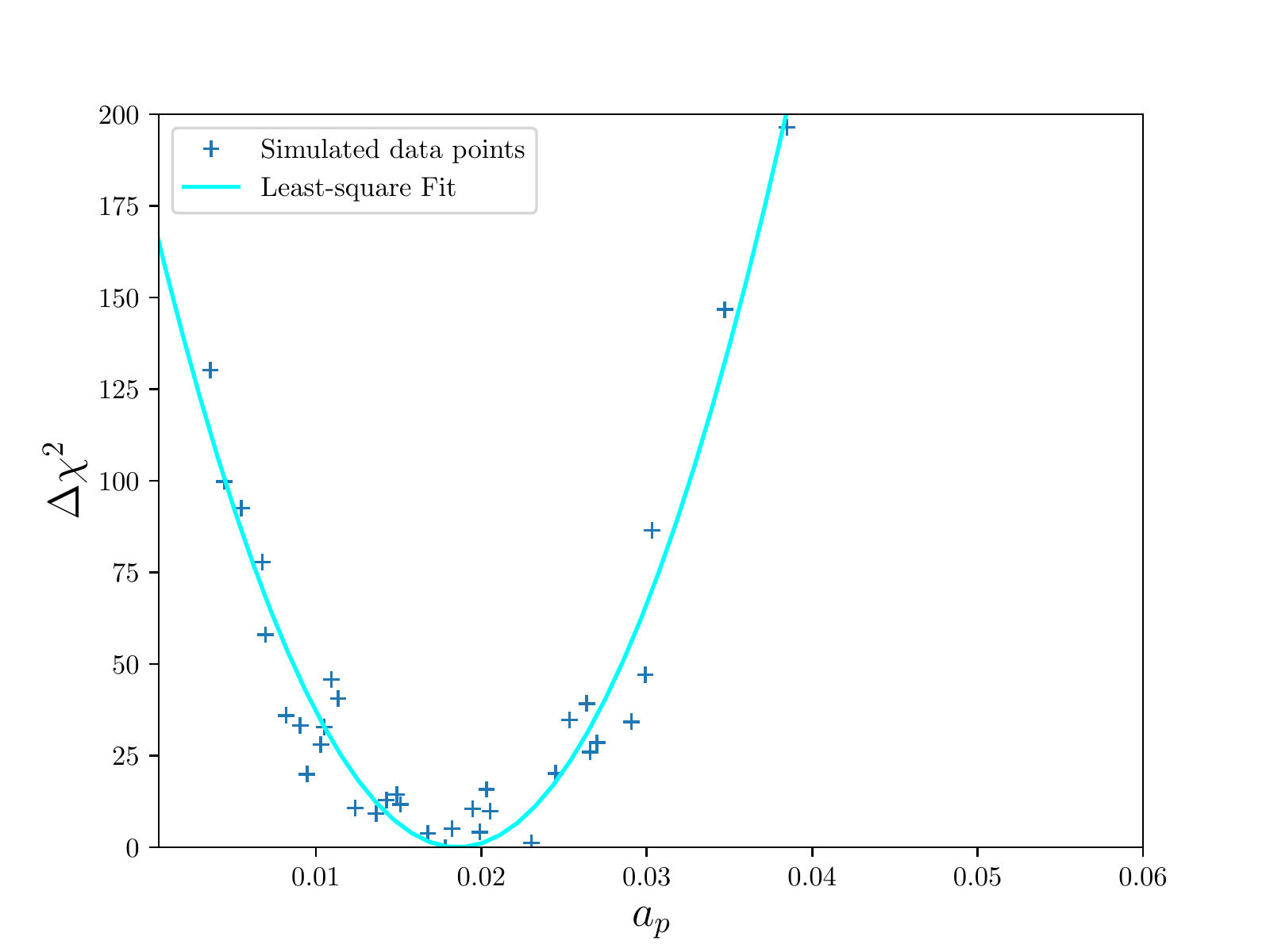}
\end{subfigure}\hspace*{\fill}
\begin{subfigure}{0.49\textwidth}
\includegraphics[width=\textwidth]{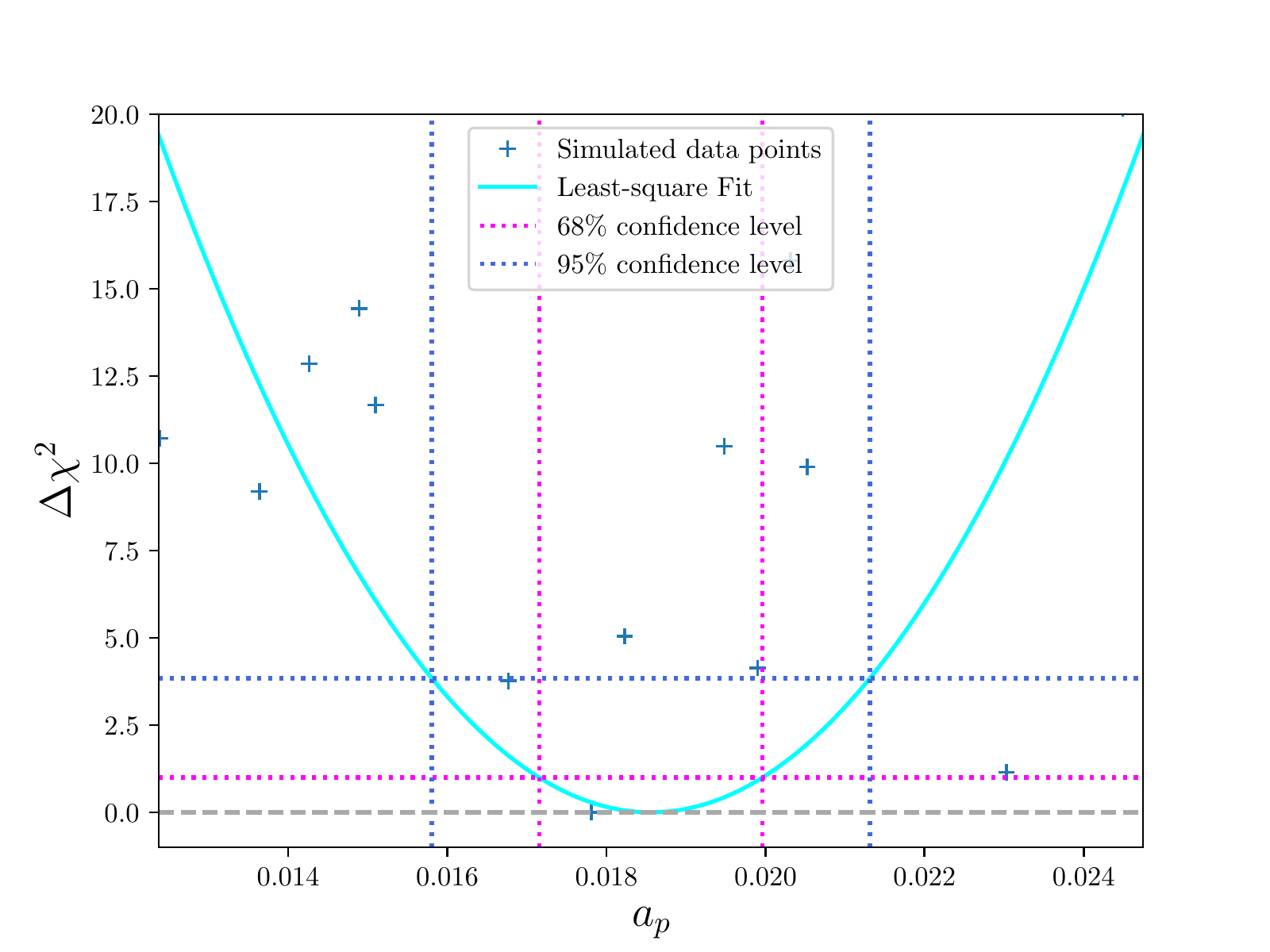}
\end{subfigure}
\caption{\protect{\label{fig_67_2}}Chi-square co-profile w.r.t. 
HiLLiPOP+lowTEB+lensing likelihood of the parameter $a_{p} = \frac{1}{z_p +1}$ together with the least-square fit 
to a parabola for $\text{SU(2)}_\CMB$ (left panel). Zoomed-in co-profile with intersections at $\Delta \chi^2 = 1, 3.84$ corresponding to 68\% and 95\% confidence levels (right panel). Central value and error range (68\% confidence) are quoted in the first column of Tab.\,\ref{tab:CosmParamFinal}.}
\end{figure*}

\noindent The utilised likelihood functions, lowTEB, HiLLiPOP, and lensing, are introduced in \cite{Adghanim2016}. Omitting the lowTEB likelihood does not influence the fit results in any essential way. Prior ranges used in 
the fits are quoted in Tab.\,\ref{tab:prior} for all free parameters. In our present 
investigation we fix $N_{\rm eff}$ to its central $\Lambda$CDM value: $N_{\rm eff}=3.046$. This is motivated by the observations discussed in Sec.\,\ref{SHBDtheta}. 
\begin{table}
\centering
\caption{Prior ranges for the free parameters. Notice that fit 
results usually do not touch the boundaries of these ranges, see Tab.\,\ref{tab:CosmParamFinal} and Fig.\,\ref{H0scatter}. }
\setlength\extrarowheight{3pt}
\label{tab:prior}
 \begin{tabularx}{0.9\columnwidth}{X c} 
 \hline\hline
Parameter & Range \\ [0.5ex] 
\hline
$\omega_\text{b,0}$\dotfill & $[0.014 \, \ldots \, 0.027]$  \\ 
$\omega_\text{pdm,0}$\dotfill & $[0.03 \, \ldots \, 0.14]$ \\ 
$\omega_\text{edm,0}$\dotfill & $[0.01 \, \ldots \, 0.1]$   \\ 
$100\, \theta_*$\dotfill & $[0.96 \, \ldots \, 1.07]$ \\
$\tau_\text{re}$\dotfill & $[0.01 \, \ldots \, 0.20]$   \\
$\ln(10^{10}\,A_s)$\dotfill & $[1.5 \, \ldots \, 3.2]$   \\
$n_s$\dotfill & $[0.4 \, \ldots \, 1.1]$   \\ 
$z_p$\dotfill & $[17.66 \, \ldots \, 2001]$  \\ 
$\beta$\dotfill & $[0 \, \ldots \, 1]$  \\ [0.5ex]
 \hline 
\end{tabularx}
\end{table}
In Tab.\,\ref{tab:CosmParamFinal} we quote our best-fit values for cosmological parameters of $\text{SU(2)}_\CMB$ and $\text{SU(2)}_\CMB$+$V_\pm$ in comparison with those of $\Lambda$CDM. Central values and error estimates for $\text{SU(2)}_\CMB$ were obtained in \cite{MADaniel} using profile likelihoods \cite{CAMEL}, 
see Fig.\,\ref{fig_67_1} and Fig.\,\ref{fig_67_2} for further explanations. 

Since the co-profiling procedure is computationally expensive, only the central values were computed by brute-force minimisation in case of the (physically disfavoured) $\text{SU(2)}_\CMB$+$V_\pm$ model, 
no error estimates were performed. 

Let us discuss the free parameters. The errors of $\omega_\text{pdm,0}$ and $\omega_\text{edm,0}$ are comparable to the error of $\omega_\text{cdm,0}$ in $\Lambda$CDM \cite{MADaniel}; the typical values for $\chi^2/n_{\text{dof}}$ match those 
of $\Lambda$CDM. Judging by Fig.\,\ref{fig-67-2}, the co-profile for $a_p$ (and thus for $z_p$) exhibits a well 
discernible minimum. On the other hand, the high value of $z_p\sim 980$ in the case of $\text{SU(2)}_\CMB$+$V_\pm$, see Tab.\,\ref{tab:CosmParamFinal},  can be identified as an extreme outlier in spite of convergence in the 
brute-force MINUIT minimisation (typical local minima, obtained under variations of the starting points, cluster around values $z_p\sim 50$). This could be due to $\text{SU(2)}_\CMB$+$V_\pm$ unphysically propagating temperature fluctuations in the quasi-particle sector.  

Notice the low values for the spectral index for primordial, scalar curvature perturbations $n_s$ in comparison with that of $\Lambda$CDM. The values of $\omega_\text{b,0}$ for both $\text{SU(2)}_\CMB$, $\text{SU(2)}_\CMB$+$V_\pm$ are low compared to the $\Lambda$CDM case and outside the 95\% confidence level 
of the BBN deuterium concordance range $0.021\le\omega_{b,0}\le 0.024$ \cite{Patrignani2017}. However, 
the central value for $\omega_\text{b,0}$, as suggested by the $^7$Li abundance, is significantly lower \cite{Patrignani2017}. The value of $\omega_\text{m,0}$ is higher than in $\Lambda$CDM. The tendency of $\text{SU(2)}_\CMB$ to predict higher values of $\omega_\text{cdm,0}$ was already observed in the analysis of Sec.\,\ref{SHBDtheta}. For $\text{SU(2)}_\CMB$+$V_\pm$ and for $\text{SU(2)}_\CMB$ the values of $\omega_\text{m,0}$ are contained in the 68\% and 95\% confidence intervals for a flat universe of the SNe Ia constraints in \cite{Betoule2014}, 
respectively. Finally, the values of $\omega_\text{pdm,0}$ are comparable with the one extracted with the \emph{left-flank} method in Sec.\,\ref{SHBDtheta}.

The (derived) values of $H_0$ and $z_{\rm re}$ are in good agreement with local 
observation, see \cite{Riess2018} and \cite{Becker2001}. Notice that $z_d<z_*$ in both $\text{SU(2)}_\CMB$ and $\text{SU(2)}_\CMB$+$V_\pm$ just like it is in $\Lambda$CDM\footnote{Due to an erroneous implementation 
of $R_{{\rm SU(2)}_\CMB}$ and assuming $\omega_{\rm pdm,0}=0$ in \cite{HH2017} this relation 
was found to be inverted: $z_d>z_*$.}. 
\begin{table*}
\centering
\caption{Best-fit cosmological parameters of the $\text{SU(2)}_\CMB$, $\text{SU(2)}_\CMB+V_\pm$, and the $\Lambda$CDM model. The best-fit parameters of $\Lambda$CDM together with their 68\% confidence intervals are taken from \protect\cite{Adghanim2016}, employing the TT,TE,EE+lowP+lensing likelihood. For $\text{SU(2)}_\CMB$, $\text{SU(2)}_\CMB+V_\pm$ we use the HiLLiPOP+lowTEB+lensing likelihood as defined in 
\protect\cite{Adghanim2016}  (lowP and lowTEB are pixel-based likelihoods). The upper section of the table quotes free parameter values, the central 
section states the values of derived parameters, and the lower section the associated $\chi^2$ and $\chi^2/n_{\text{dof}}$ values for the low-$l$ ($2\le l\le 29$) and the high-$l$ ($30\le l$) 
spectra with respective subscripts $ll$ and $hl$. Here $n_{\text{dof}}$ is defined as the difference 
between the number of degrees of freedom and the number of free parameters. 
The computations of $z_*$ and $z_{d}$ were performed according to the conventional criterion of Eq.\,(\ref{cond:def:methodOneStar}) or its baryon-drag version. Practically, there is no dependence of other 
best-fit cosmological parameters on the value of $\alpha$, defined in Eq.\,(\ref{inidelta}), whereas this is not the case for $\beta$, defined in Eq.\,(\ref{initheta}). 
For $\text{SU(2)}_\CMB$ central values are identified with the minima of co-profiles \protect\cite{MADaniel}.
Errors correspond to 68\%-confidence levels, see Fig.\,\ref{fig_67_1} and 
Fig.\,\ref{fig_67_2} for the cases $\omega_\text{b,0}$ and $a_p=\frac{1}{z_p+1}$, respectively. For the physically less motivated case of $\text{SU(2)}_\CMB+V_\pm$, see Sec.~\ref{sec:boltzmannstuff}, we only consider the output of brute-force minimisation using MINUIT (contained in CAMEL) and refrain from performing the computationally expensive co-profiling procedure \protect\cite{MADaniel}.
The value of $\sigma_8$ was only computed for 
$\text{SU(2)}_\CMB$ since its determination for $\text{SU(2)}_\CMB+V_\pm$ would have 
necessitated further modifications of CLASS. The central values associate with $\chi^2=\chi^2_{ll}+\chi^2_{hl}$ (best fit) as quoted below. }
\setlength\extrarowheight{3pt}
\label{tab:CosmParamFinal}
 \begin{tabularx}{1.50\columnwidth}{X c c c} 
 \hline\hline
Parameter & $\text{SU(2)}_\CMB$ & $\text{SU(2)}_\CMB$+$V_\pm$ & $\Lambda$CDM  \\ [0.5ex] 
\hline
$\omega_\text{b,0}$\dotfill & $0.0173 \pm 0.0002$ & 0.0167 & $0.0225 \pm 0.00016$ \\ 
$\omega_\text{pdm,0}$\dotfill & $0.113 \pm 0.002$ & 0.106 & $-$ \\ 
$\omega_\text{edm,0}$\dotfill & $0.0771 \pm 0.0012$  & 0.0685 & $-$\\ 
$100\, \theta_*$\dotfill & $1.0418 \pm 0.0022$ & 1.0418 & $1.0408 \pm 0.00032$  \\
$\tau_\text{re}$\dotfill & $0.02632 \pm 0.00218$ & 0.0166 & $0.079 \pm 0.017$ \\
$\ln(10^{10}A_s)$\dotfill & $2.858 \pm 0.009$ & 2.799 & $3.094 \pm 0.034$ \\
$n_s$\dotfill & $0.7261 \pm 0.0058$  & 0.7022  &$0.9645 \pm 0.0049$ \\ 
$z_p$\dotfill & $52.88\pm 4.06$ & 981.45 &$-$\\ [0.5ex]
$\beta$\dotfill & $0.811\pm0.058$& 0.72 &$-$\\ [0.5ex]
 \hline
$H_0/${\tiny{km\,s$^{-1}$Mpc$^{-1}$}} \dotfill & $74.24 \pm 1.46$ & 73.41 &$67.27\pm 0.66$ \\ [0.5ex]
$z_\text{re}$\dotfill & $6.23 ^{+0.41}_{-0.42}$ & 4.41 & $10^{+1.7}_{-1.5}$ \\
$z_*$\dotfill & $1715.19 \pm 0.19$ & 1716.55 & $1090.06 \pm 0.30$ \\
$z_{d}$\dotfill & $1640.87 \pm 0.27$  & 1639.49 & $1059.65 \pm 0.31$ \\
$\omega_\text{cdm,0}$\dotfill & $0.1901 \pm 0.0023$ & 0.1745 & $0.1198\pm 0.0015 $\\ 
$\Omega_\Lambda$\dotfill & $0.616 \pm 0.006$ & 0.645 & $0.6844\pm 0.0091$ \\
$\Omega_{\text{m,0}}$\dotfill & $0.384 \pm 0.006$ & 0.355 & $0.3156\pm 0.0091$ \\
$\sigma_8$\dotfill & $0.709 \pm 0.020$ & $-$ & $0.8150\pm 0.0087$ \\
Age$/$Gyr\dotfill & $11.91 \pm 0.10$ & 12.25 & $13.799 \pm 0.021$ \\ [0.5ex]
\hline
$\chi^2_{ll}$\dotfill & 10640 & 10585.6 & 10495 \\
$n_{\text{dof},ll}$\dotfill & 9207 &  9207 & 9210\\
$\frac{\chi^2_{ll}}{n_{\text{dof},ll}} $\dotfill & 1.156 & 1.150 & 1.140\\
$\chi^2_{hl}$\dotfill & 10552.6 & 10354.5 & 9951.47 \\
$n_{\text{dof},hl}$\dotfill & 9547 & 9547 & 9550\\
$\frac{\chi^2_{hl}}{n_{\text{dof},hl}}$ \dotfill & 1.105 & 1.085 & 1.042\\ [0.5ex]
\hline
\end{tabularx}
\end{table*}

\begin{figure*}
\centering
\includegraphics[width=\textwidth]{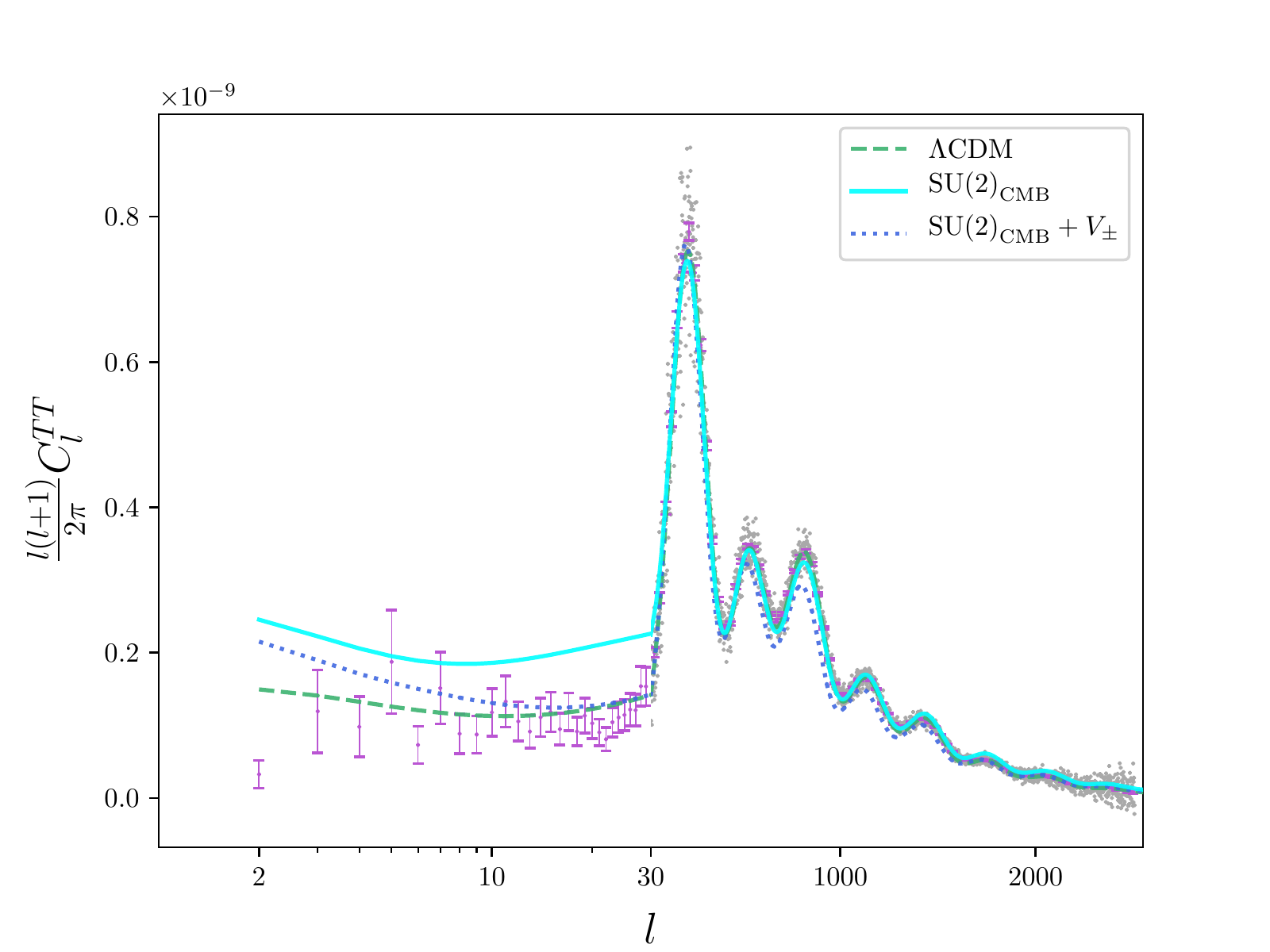}
\caption{\protect{\label{TT}} 
Normalised power spectra of TT correlator for best-fit parameter values 
quoted in Tab.\,\ref{tab:CosmParamFinal}: Dashed, dotted, and solid lines represent $\Lambda$CDM, $\text{SU(2)}_\CMB$+$V_\pm$, and $\text{SU(2)}_\CMB$, respectively. For $l \leq 29$ the 2015 Planck data points are unbinned and carry error bars, for $l\ge 30$ grey points represent unbinned spectral power.   
}
\end{figure*}

Let us now discuss the angular power spectra. In Fig.\,\ref{TT} the normalised spectra of the TT correlator are shown for best-fit $\text{SU(2)}_\CMB$, $\text{SU(2)}_\CMB$+$V_\pm$, and $\Lambda$CDM. In spite of $\chi^2$ being slightly smaller in $\text{SU(2)}_\CMB$+$V_\pm$ compared to $\text{SU(2)}_\CMB$, a visual inspection of the first three peaks suggests that the latter model fits the data better than the former one. 

At low $l$ both $\text{SU(2)}_\CMB$ and $\text{SU(2)}_\CMB$+$V_\pm$ exhibit excesses in spectral power. This is induced by the small values of the spectral index $n_s$, required to fit the high-$l$ region. One should keep in mind, however, that screening effects at low $z$, neglected in the present work, break statistical isotropy and suppress the TT correlator at large angles (for a review see \cite{Hofmann2013}, for original work see \cite{SzopaHofmann2007,LudescherHofmann2008}). Assuming spherical symmetry, these effects potentially account for the discrepancy between the observed and the Doppler inferred CMB dipole. Therefore, after a relaxation of this assumption, $\text{SU(2)}_\CMB$ induced screening is likely to substantially suppress the low-$l$ multipoles. Since the extraction of the $C_l$'s from a given temperature map {\sl assumes} statistical isotropy, they cease to be meaningful. A quantitative assessment of a cosmological model in terms of data on the TT correlator at large angles thus requires alternative statistics \cite{Schwarz2016}. We leave this complex investigation in $\text{SU(2)}_\CMB$ to the future.

\begin{figure}
\centering
\includegraphics[width=\columnwidth]{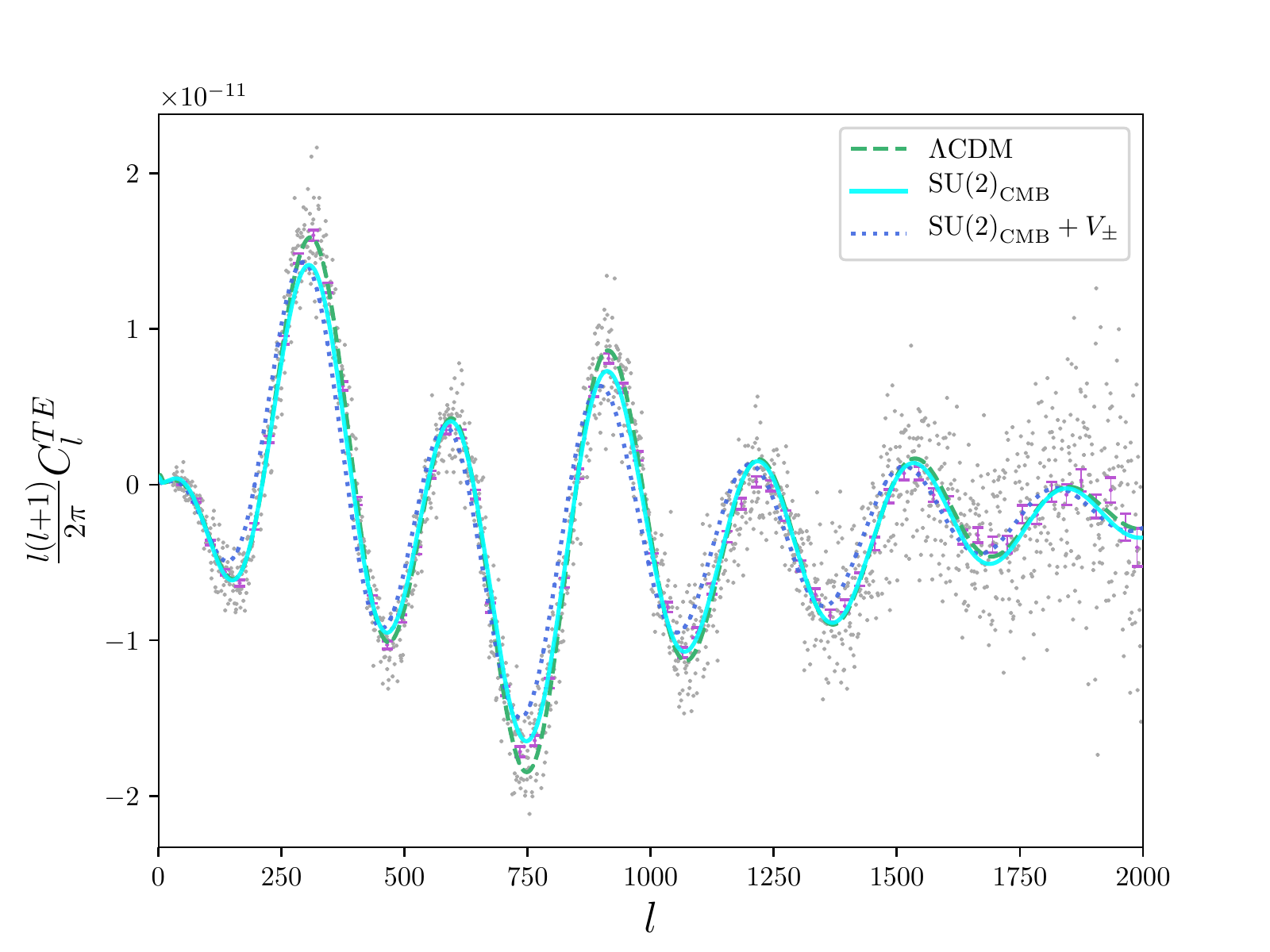}
\caption{\protect{\label{TE}}
Normalised power spectra of TE cross correlator for best-fit parameter values 
quoted in Tab.\,\ref{tab:CosmParamFinal}: Dashed, dotted, and solid lines represent $\Lambda$CDM, $\text{SU(2)}_\CMB$+$V_\pm$, and $\text{SU(2)}_\CMB$, respectively. The 2015 Planck data points are either unbinned without error bars (grey) or binned with error bars (blue).  
}
\end{figure}

\begin{figure}
\centering
\includegraphics[width=\columnwidth]{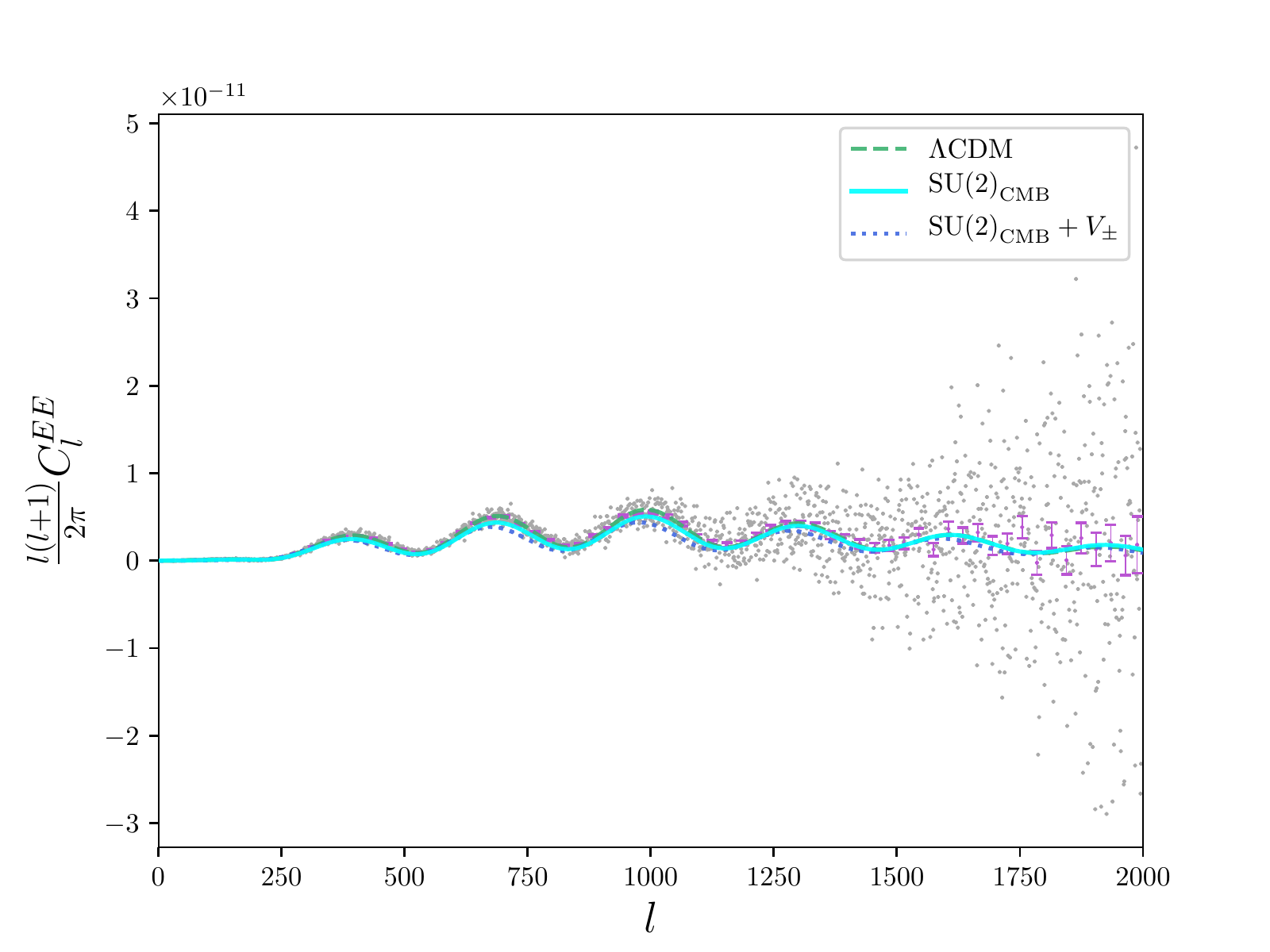}
\caption{\protect{\label{EE}}
Normalised power spectra of EE cross correlator for best-fit parameter values 
quoted in Tab.\,\ref{tab:CosmParamFinal}: Dashed, dotted, and solid lines represent $\Lambda$CDM, $\text{SU(2)}_\CMB$+$V_\pm$, and $\text{SU(2)}_\CMB$, respectively. The 2015 Planck data points are either unbinned without error bars (grey) or binned with error bars (blue).  
}
\end{figure}

In Figs.\,\ref{TE} and \ref{EE} the spectra of the TE and EE cross correlators are shown: they do not differentiate the cosmological models considered and agree with the data.

\begin{landscape}
\begin{figure}
\centering
\includegraphics[width=1.25\textwidth]{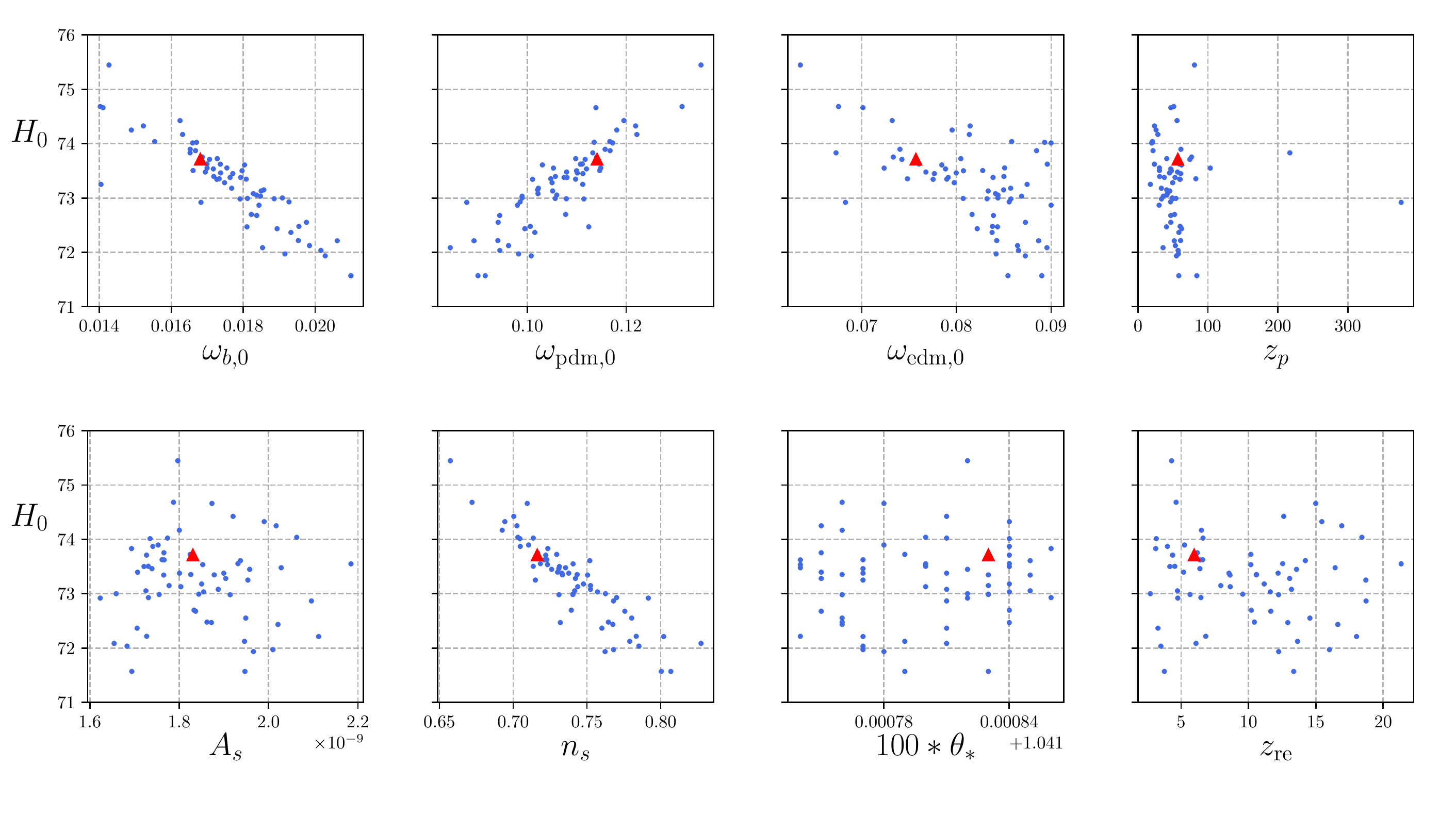}
\caption{\protect{\label{H0scatter}}
Scatter plots of $H_0$ for fitted parameter values in case of $\text{SU(2)}_\CMB$, requiring that $\chi^2=\chi^2_{hl}+\chi^2_{ll}<21700$ in brute-force minimisation using MINUIT (best-fit: $\chi^2=21192.6$). 
Dots refer to local minima, best-fit parameter values are indicated by a solid red triangle.}
\end{figure}
\end{landscape}

Finally, we look at the value of $H_0$ in dependence of all parameter values, requiring $\chi^2 \leq 21700$, see Fig.\,\ref{H0scatter}. There is a conspicuous clustering around the local value of $H_0 \sim\,73.5$\,km\,s$^{-1}$Mpc$^{-1}$ \cite{Riess2018,Bonvin2017}; there is a less pronounced clustering around the low value of $z_\text{re}\sim\, 6$ as observed in \cite{Becker2001}.

\section{Summary and discussion}
\label{SD}

In the present work a cosmological model was analysed, describing a spatially flat universe and assuming the CMB to associate with an SU(2) Yang-Mills theory. The scale of this theory is fixed by the excess in line temperature at low frequencies observed by terrestrial and balloon-borne experiments, see \cite{Fixsen2011, Hofmann2009} and references therein. Driven by an unconventional temperature-redshift relation this model requires a modified dark sector which transforms into the familiar $\Lambda$CDM model at low redshifts. With decreasing redshift, this is facilitated by an instantaneous depercolation process, characterised by the emergence of a certain amount of dark matter. The model also exhibits a neutrino sector which is subject to a modified conversion between CMB and neutrino temperatures. 

In a first step, we have investigated how the constraints of the standard ruler (co-moving sound horizon at baryon drag), inferred from $H_0$'s local value \cite{Bernal2016, Bonvin2017, Riess2018}, and of the angular size of the sound horizon at photon decoupling fix two out of the six parameters governing the cosmological expansion. 
Compared to $\Lambda$CDM, we obtain a $\sim$20\% excess of today's physical total dark-matter density and a fraction of primordial to today's total dark matter ranging from 0.5 to 0.7, subject to $H_0\sim 73.5\,$\,km\,s$^{-1}$Mpc$^{-1}$ \cite{Riess2018}. Other parameters of this model in addition to 
those introduced by linear perturbations can be determined by fits to the CMB angular power spectra TT, TE, and EE of the 2015 Planck data release \cite{Ade2016}. 

In a second step, we have considered two scenarios of accounting for the temperature fluctuations associated with the massive quasi-particle excitations $V_\pm$ in the deconfining SU(2) Yang-Mills plasma. Neglecting 
$V_\pm$ in the perturbation equations is favoured over taking them into account in view of TT data as well as theoretically~\cite{Hofmann2017}. The impact of a change in scenario is negligible in the 
predictions of TE and EE. Presently, we have neglected any radiative, screening effects in the photon dispersion law, which play out at low redshifts and influence low multipoles, see \cite{Hofmann2013} and references therein.

Spectra are computed with a modified version of CLASS (Cosmic Linear Anisotropy Solving System \cite{CLASS}). Employing typical likelihood functions of the Planck data-analysis campaign \cite{Adghanim2016} to fit these spectra, we confirm the tendency of an excess in total dark matter, and we find a red-tilted spectrum of adiabatic primordial curvature perturbations ($n_s \sim 0.7$) as well as low values for baryon density $\omega_{b,0}$ and reionisation depth. The low value of $n_s$ contradicts the standard scenario of inflationary, scale-invariant density perturbations, generated by a single slowly rolling inflaton field. In particular, our value $n_s \sim 0.7$ violates the slow-roll approximation, see, e.g., \cite{Mukanov1997} and \cite{Martin2000}, since the leading-order term for $n_s$ as a function of the slow-roll parameter $\epsilon\sim 0.15$ predicts a next-to-leading order corrections of about 20\%, and an expansion beyond next-to-leading order is inconsistent \cite{Mukanov1997}. One therefore would have to apply the exact method, see \cite{Mukanov1997} and references therein, to predict the-no-longer scale invariant spectrum of adiabatic curvature perturbations from the inflaton dynamics. The epoch of inflation thus would correspond to an inflaton that appreciably rolls down its potential, implying sizable variations of the Hubble parameter. 

The here-obtained excess of spectral TT power at low multipoles may reduce or completely be compensated 
by consideration of radiatively induced screening effects \cite{Hofmann2013,SzopaHofmann2007,LudescherHofmann2008}. On the other hand, the associated breaking of statistical isotropy at low redshift casts doubts on the meaningfulness of the $C_l$'s at low $l$. Rather, alternative statistics should be used to unravel the physics at large angles in the CMB~\cite{Schwarz2016}. 

There is tension of $\omega_{b,0}$ with the observed deuterium abundance, matched to the BBN prediction \cite{Patrignani2017}. Presently, we interpret this as a shortcoming of our cosmological model. Yet, future reconsideration of certain assumptions in BBN together with the here-obtained low value for $\omega_{b,0}$ could contribute to a resolution of the $^7$Li puzzle. This is speculative, however, and should be solidly investigated. It is instructive, however, to compare our fit results with those of direct baryon censuses. The results discussed in \cite{Shull2011} suggest that baryon density mainly resides in the photoionised Lyman forest and in the warm-hot intergalactic medium (WHIM) while the collapsed phase yields a smaller correction. In total, this only amounts to about 70\% of the baryons predicted by the coincidence in deuterium via BBN and the $\Lambda$CDM fit to the Planck data. (There is, however, a problem with Li.) Interestingly, this value ($0.7\times \Omega_{\rm b,0,\Lambda CDM}=0.0348$ or $0.7\times \omega_{\rm b,0,\Lambda CDM}=0.0158$) is close to our fit results (first column of Table 1): $\Omega_{\rm b,0, SU(2)_{CMB}}=0.0314$ or $\omega_{\rm b,0,SU(2)_{CMB}}=0.0173$. There is, however, a recently announced strong claim in \cite{Nicastro2018} that the missing 30\% of baryons have been found in the hotter phases of WHIM by observing O VII absorbers in the X-ray spectra of quasars. These authors base their results on theoretical simulations of the hot, metal-rich filaments as well as the missing temporal structure of the signal and a missing associated cold absorption. This appears to exclude the possibility that the quasars'/blazars' outflows or the host galaxies' interstellar media contribute to the column densities. On the other hand, these authors declare that the sources are near the absorbing systems, and that therefore it cannot be excluded that column densities are contaminated by source outflows. To clarify the situation affirmatively, independent obervations with a clear separation of source and absorbing system should be performed.
Finally, the redshift of depercolation is favoured to be 30, that is, well within the dark ages.

As for derived parameters, our best fits favour central values for $H_0$ in good agreement with the extractions from local cosmology \cite{Bonvin2017, Riess2016, Riess2018}. The total matter density is about 20\% higher than that of $\Lambda$CDM. Although this is at some tension with present bounds inferred from SNe Ia Hubble diagrams~\cite{Betoule2014}, it is not excluded. The best-fit values of the redshift $z_\text{re}$ for the reionisation of the universe favour the value $z_\text{re} \sim 6$ extracted from the Gunn-Peterson trough~\cite{Becker2001}. 

An urgent field of research is a quantification of the screening effects for the photon, concerning a possible explanation of the suppression of the TT correlator at large angles, various power asymmetries, and the violation of Gaussianity about the CMB cold spot~\cite{Schwarz2016,Hofmann2013,Vielva2010}.
Moreover, an analysis of alignments and power of the low multipoles, as induced by such screening effects, should be performed along the lines of~\cite{Schwarz2016}.
Also, a ``microscopic'' understanding of the depercolation process involving the vortices of a Planck-scale axion field~\cite{Frieman1995,Neubert} is mandatory to validate/improve/falsify our present parametrisation of the dark sector.

\section*{Acknowledgements} 

We would like to thank Thomas Schwetz-Mangold for continuing, helpful discussions and 
Oliver Fischer for useful conversations on inflationary models. 

%
%
%

\label{lastpage}
\end{document}